\documentclass[prd,amssymb,amsmath,amsfonts,superscriptaddress,nofootinbib,reprint,showpacs, longbibliography]{revtex4-1}

\usepackage{graphicx}
\usepackage{lmodern}
\usepackage{amsmath,amssymb}
\usepackage{mathrsfs}
\usepackage{amsfonts}
\usepackage[utf8]{inputenc}
\usepackage{url}
\usepackage[colorlinks]{hyperref}
\usepackage{xcolor}
\usepackage{multirow}
\usepackage[normalem]{ulem}
\usepackage{orcidlink}
\usepackage{placeins}
\usepackage{mathtools}
\usepackage[shortlabels]{enumitem}

\definecolor{dodgerblue}{HTML}{1E90FF}
\definecolor{viennared}{HTML}{DA0A14}
\definecolor{ctorange}{HTML}{FF6C0C}
\definecolor{wales}{HTML}{ff0038}
\definecolor{benettongreen}{HTML}{009421}
\definecolor{ferrarired}{HTML}{ff2800}
\definecolor{austriawienpurple}{HTML}{441678}

\DeclareFontFamily{OT1}{pzc}{}
\DeclareFontShape{OT1}{pzc}{m}{it}{<-> s * [1.10] pzcmi7t}{}
\DeclareMathAlphabet{\mathpzc}{OT1}{pzc}{m}{it}

\hypersetup{
     colorlinks=true,
     linkcolor=dodgerblue,
     filecolor=dodgerblue,
     citecolor = dodgerblue,      
     urlcolor=dodgerblue,
     }

\newcommand{\CIT}{TAPIR, California Institute of Technology, Pasadena, CA 91125, USA}
\newcommand{\LIGOLab}{LIGO Laboratory, California Institute of Technology, Pasadena, California 91125, USA}
\newcommand{\UMassD}{Department of Mathematics,
    Center for Scientific Computing and Data Science Research,
    University of Massachusetts, Dartmouth, MA 02747, USA}

\newcommand{\tensorflow}{\texttt{Tensorflow}}
\newcommand{\numpy}{\texttt{NumPy}{ }}
\newcommand{\jax}{\texttt{JAX}{ }}

\newcommand{\optuna}{\texttt{Optuna}{ }}
\newcommand{\scikitlearn}{\texttt{scikit-learn}{ }}
\newcommand{\gwbonsai}{\texttt{gwbonsai}{ }}

\newcommand{\Y}{\textbf{Y}{ }}
\newcommand{\X}{\textbf{X}{ }}
\newcommand{\Remnant}{\texttt{NRSur7dq4Remnant}{}}
\newcommand{\RemnantNN}{\texttt{NRSur7dq4Remnant\_NN}{}}
\newcommand{\Hset}{$\mathpzc{H}$-set}
\newcommand{\Tset}{$\mathcal{T}$-set}
\newcommand{\Vset}{$\mathcal{V}$-set}
\newcommand{\Sset}{$\mathcal{S}$-set}

\DeclarePairedDelimiterX{\norm}[1]{\lVert}{\rVert}{#1}

\begin{document}

\title{Optimizing Neural Network Surrogate Models: Application to Black Hole Merger Remnants}

\author{Lucy M. Thomas \orcidlink{0000-0003-3271-6436}}
\email{lmthomas@caltech.edu}
\affiliation{\CIT}
\affiliation{\LIGOLab}
\author{Katerina Chatziioannou \orcidlink{0000-0002-5833-413X}}
\email{kchatziioannou@caltech.edu}
\affiliation{\CIT}
\affiliation{\LIGOLab}
\author{Vijay Varma}
\affiliation{\UMassD}
\author{Scott E. Field \orcidlink{0000-0002-6037-3277}}
\affiliation{\UMassD}

\date{\today}

\begin{abstract}

Surrogate models of numerical relativity simulations of merging black holes provide the most accurate tools for gravitational-wave data analysis.
Neural network-based surrogates promise evaluation speedups, but their accuracy relies on (often obscure) tuning of settings such as the network architecture, hyperparameters, and the size of the training dataset.
We propose a systematic optimization strategy that formalizes setting choices and motivates the amount of training data required.
We apply this strategy on \Remnant, an existing surrogate model for the properties of the remnant of generically precessing binary black hole mergers and construct a neural network version, which we label \RemnantNN. 
The systematic optimization strategy results in a new surrogate model with comparable accuracy, and provides insights into the meaning and role of the various network settings and hyperparameters as well as the structure of the physical process.    
Moreover, \RemnantNN~ results in evaluation speedups of up to $8$ times on a single CPU and a further improvement of $2,000$ times when evaluated in batches on a GPU.
To determine the training set size, we propose an iterative enrichment strategy that efficiently samples the parameter space using much smaller training sets than naive sampling. 
\RemnantNN~requires ${\cal{O}}(10^4)$ training data, so neural network-based surrogates are ideal for speeding-up models that support such large training datasets, but at the moment cannot directly be applied to numerical relativity catalogs that are ${\cal{O}}(10^3)$ in size. 
The optimization strategy is available through the \gwbonsai package.
\end{abstract}

\maketitle 

\section{Introduction}
\label{sec:Introduction}

Binary black hole (BBH) mergers account for most gravitational waves detected so far \cite{LIGOScientific:2018mvr,LIGOScientific:2020ibl,LIGOScientific:2021usb,KAGRA:2021vkt,Nitz:2018imz,Nitz:2020oeq,Nitz:2021uxj,Nitz:2021zwj,Venumadhav:2019tad,Venumadhav:2019lyq,Olsen:2022pin,Mehta:2023zlk} by the LIGO \cite{LIGOScientific:2014pky}, Virgo \cite{VIRGO:2014yos} and KAGRA \cite{KAGRA:2020tym} detectors. 
Measuring and interpreting the system properties relies on models for the observed signal \cite{Cutler:1994ys,Veitch:2014wba,Ashton:2018jfp}. 
Numerical relativity (NR) simulations of the binary dynamics and gravitational emission in full General Relativity provide the only accurate description of the BBH merger, including the late inspiral and merger signals as well as the remnant properties. 
However, the high computational cost and relatively sparse parameter space coverage \cite{Boyle:2019kee,Healy:2022wdn,Ferguson:2023vta} makes it infeasible to rely solely on NR simulations, as data analysis requires millions of model evaluations. 
Approximate semi-analytical models for waveforms \cite{Ajith:2009bn,Santamaria:2010yb,Pratten:2020fqn, Garcia-Quiros:2020qpx,Pratten:2020ceb,Buonanno:1998gg,Pompili:2023tna,Khalil:2023kep,Ramos-Buades:2023ehm,Nagar:2023zxh,Gamba:2024cvy,Colleoni:2024knd} and remnant properties \cite{Campanelli:2007cga,Hofmann:2016yih,Barausse:2012qz,Jimenez-Forteza:2016oae,Healy:2016lce,Healy:2014yta,Gonzalez:2006md,Campanelli:2007ew,Lousto:2007db,Lousto:2012su,Lousto:2012gt,Gerosa:2016sys,Healy:2018swt,Herrmann:2007ex,Barausse:2012qz,Rezzolla:2007rz,Rezzolla:2007rd,Kesden:2008ga,Tichy:2008du,Barausse:2009uz,Zlochower:2015wga} that combine analytical insights with fits to NR have therefore been developed.

An alternative approach involves direct surrogate models to the NR simulations, which are fast and accurate approximations to a more computationally intensive underlying model \cite{Field:2013cfa}.
NR surrogates match simulations more accurately than semi-analytical models, but they are limited by the parameter space coverage and availability of NR catalogs.
Existing NR and NR-hybrid surrogate models target both waveforms and remnant properties~\cite{Blackman:2015pia, Blackman:2017dfb, Varma:2018mmi,Varma:2018aht,Varma:2019csw, Islam:2021mha, Islam:2022laz,Yoo:2022erv,Yoo:2023spi}. 
Surrogate construction optionally first reduces the training data to sufficiently sparse, smoothly-varying components,
and then fits these components across the parameter
space~\cite{Field:2013cfa}. 
Fitting methods include polynomials~\cite{Field:2013cfa,Blackman:2017dfb},
splines~\cite{Galley:2016mvy,Purrer:2015tud}, Gaussian Process Regression
(GPR)~\cite{Varma:2018mmi,Varma:2019csw}, or more recently neural
networks~\cite{Setyawati:2019xzw,Khan:2020fso,Schmidt:2020yuu,Fragkouli:2022lpt,Thomas:2022rmc,Shi:2024age,Tiki:2024jqx,GramaxoFreitas:2024bpk,Que:2021cqo,Chua:2018woh,Tissino:2022thn}.
 
Neural networks can yield comparable accuracy to other methods while significantly reducing model evaluation time.
However, their efficacy hinges on many seemingly arbitrary choices of settings, such as the neural network architecture or its numerous hyperparameters, which are not straightforwardly physically motivated. 
The impact of these choices is typically not clear until after the expensive training process has been completed and the network's accuracy can be assessed.
The high cost of full network training makes exploring multiple choices for the architecture and hyperparameters cumbersome. 

Hyperparameter optimization focuses on finding the best hyperparameters, often within a high-dimensional space, to maximize model performance. Common techniques include grid and random search (simple but often inefficient), Bayesian optimization (uses probabilistic models to balance exploration and exploitation of the search space), and evolutionary algorithms (mimic natural selection through meta-heuristic strategies). Many of these methods are implemented in flexible hyperparameter optimization tools such as Optuna, Ray Tune, Keras Tuner, Auto-sklearn, DEAP, and HyperOpt, making them suitable for a wide range of problems.

Solving the hyperparameter optimization problem is increasingly important in gravitational wave astrophysics. Gradient-free genetic algorithms, for instance, have been shown to refine deep matched filtering architectures for greater accuracy and compactness~\cite{Deighan:2020gtp}. In waveform modeling studies using neural networks~\cite{Setyawati:2019xzw,Khan:2020fso,Schmidt:2020yuu,Fragkouli:2022lpt,Thomas:2022rmc,Shi:2024age,Tiki:2024jqx,GramaxoFreitas:2024bpk,Que:2021cqo,Chua:2018woh,Tissino:2022thn}, most hyperparameters are set heuristically or through basic random/grid search. Notable exceptions include neural-network-based waveform models for aligned-spin systems~\cite{Tissino:2022thn,GramaxoFreitas:2024bpk} and precessing systems approximated as an aligned-spin system in the coprecessing frame~\cite{Thomas:2022rmc}. To date, no work has systematically explored hyperparameter or training-set optimization for fully generic precessing BBH systems.

In this work, we propose a framework that systematically optimizes the training settings, including the network architecture, network hyperparameters, and the amount of training data.
As an application of this framework, we consider the remnant surrogate \Remnant
~\cite{Varma:2019csw} which calculates the merger remnant properties (final black hole (BH)
mass, spin, and recoil velocity) from the BBH properties (initial BH mass ratio
and spins). 
The remnant mass and spin determine the ringdown mode content~\cite{PhysRevD.1.2870,Press:1971wr,PhysRevD.5.2419,PhysRevD.5.2439,Berti:2009kk,Teukolsky:2014vca}, while the recoil velocity determines whether the remnant is ejected from its host environment~\cite{Merritt:2004xa,Campanelli:2007cga,Gonzalez:2007hi,Gerosa:2014gja,Varma:2022pld,Varma:2020nbm}.
For maximum flexibility, we do not work directly with NR data but instead use \Remnant~to make training data that we then fit with a neural network. 

We address three key elements of efficient and accurate neural network-based surrogates.
\begin{enumerate}
    \item  We explore the interplay between neural network architecture and hyperparameters values toward optimal accuracy. 
    Using optimization techniques, we study the role of various settings, in the process clarifying their role during network training.
    \item We examine whether splitting the output parameters among multiple networks allows for specialization in training and improved accuracy without overfitting.
    We find no accuracy gain from considering parameter subsets separately. 
    What is more, the full 7-dimensional network achieves high accuracy with relatively less aggressive settings than some subset networks. 
    \item We investigate the relation between the size of the training data and model accuracy, determining the minimum amount of data necessary to achieve high-precision predictions without unnecessary computational overhead.
    We find that ${\cal{O}}(10^4)$ training data points are required, compared to the ${\cal{O}}(10^3)$ NR points \Remnant~was fit on initially. 
\end{enumerate}

Using the optimal settings, we train a new neural network-based surrogate model, \RemnantNN.
The accuracy of the new model compared to the underlying model it was trained on is comparable or better than the underlying model's accuracy to its underlying NR data.
The upper 95th percentile error over all $7$ output parameters between \RemnantNN~and \Remnant~is $1.1\times10^{-2}$ while the same quantity between \Remnant~and the original NR data is $4.1\times10^{-2}$.
Moreover, \RemnantNN~unlocks significant computational improvements.
When evaluated on a CPU, we obtain speedups of up to $8$ times (comparing the median time across the fastest implementation of each model), while batch evaluation on a GPU results in a further speedup of $2,000$.
The optimization framework is model-agnostic, it is therefore applicable to other surrogates which use interpolation, including waveform surrogates.

We release the optimization code as \gwbonsai~\cite{gwbonsai}. 
The package uses optimization procedures from \texttt{Optuna} \cite{optuna_2019} (or alternatively \texttt{Hyperopt} \cite{Hyperopt:2013byc}), and allows users to automatically optimize network settings as part of the surrogate model construction. 
It can utilize either a \tensorflow \cite{tensorflow2015-whitepaper} or \texttt{PyTorch} \cite{Paszke:2019xhz} backend for network training, and produces both GPU-evaluable results designed for batch evaluation and a \numpy \cite{harris2020array} (or \jax \cite{jax2018github}) implementation suited for evaluation on CPUs. 

The paper is organized as follows. 
In Sec.~\ref{sec:ModelStructureandDatasets} we provide an overview of the construction and training of \RemnantNN. 
In Sec.~\ref{sec:Hyperparameteroptimization} we develop infrastructure to optimize the neural network architecture and hyperparameters for several subsets of output parameters. 
In Sec.~\ref{sec:TrainingDatasetSizeOptimization} we assess the optimal training dataset size and configuration, again considering several output subsets. 
In Sec.~\ref{sec:FinalModelEvaluation} we train the final optimized network and evaluate its performance in terms of accuracy, timing, and extrapolation performance, before concluding in Sec.~\ref{sec:Conclusions}.
Throughout this paper we use geometric units, $G=c=1$, unless stated otherwise.

\section{Model Structure and Datasets}
\label{sec:ModelStructureandDatasets}

We apply the network optimization framework to the remnant surrogate model \Remnant, and construct a neural network-based surrogate-of-the-surrogate, \RemnantNN. 
In this section we describe the \Remnant~ structure and the relevant datasets. 
More details can be found in Ref.~\cite{Varma:2019csw} and the supplemental material in Ref.~\cite{Varma:2018aht}.
The model represents a fit of 7 input parameters to 7 output parameters.
The input parameters are the mass ratio, $q$, and dimensional spin components of
the two pre-merger BHs, $\vec{\chi}_i$ with $i\in\{1,2\}$, at a reference time
$t=-100\,M$ before merger\footnote{Additional 
functionality allows for input spins specified at earlier
frequencies and then evolved to $100\,M$ before merger. We do not consider spin
evolution here either in the fits or accuracy tests.} in the coorbital
frame\footnote{In the coorbital frame the $z$-axis aligns with the orbital
angular momentum, and the $x$-axis is defined as the direction from the smaller
to the larger BH.} (with $M$ the binary's total mass), while the 7 output
parameters are the mass, $m_f$, spin, $\vec{\chi}_f$, and recoil velocity,
$\vec{v}_f$, of the final BH. \Remnant~was constructed through a separate GPR fit for each of its 7 output parameters, using a kernel given in Eq.~(S3) of Ref.~\cite{Varma:2018aht}.

We then construct a neural network surrogate of \Remnant.
The output parameters are the same (remnant mass, spin, and velocity).
For the input parameters, we use the mass ratio and spins in polar coordinates.
Unlike Ref.~\cite{Varma:2018aht}, we do not reparametrize the spins to quantities motivated by post-Newtonian theory, as initial investigations suggest that the polar parameterization leads to the highest accuracy. 
Following Ref.~\cite{Varma:2018aht}, we standard-scale the input and output training data such that the mean is subtracted and the standard deviation re-scaled to $1$, using \scikitlearn\cite{scikit-learn}.
Training is then based on fully-connected Multi-Layer Perceptron networks, with each layer (except the input and output ones) having the same number of neurons, as we find that this architecture is sufficiently complex for our target accuracy. 
We use \texttt{Tensorflow} for training and testing, though \gwbonsai can alternatively utilize \texttt{PyTorch}.

We construct four datasets with \Remnant: a training set with which to train the neural network, the \Tset; a validation set for use in training to avoid overfitting, the \Vset; a test set to evaluate the accuracy of the trained network, the \Sset; and a hold-out set, the \Hset. 
All sets comprise values for input parameters 
\begin{equation}
    \X=\lbrace q, \chi_{1}, \theta_1, \phi_1, \chi_{2}, \theta_2, \phi_2 \rbrace\,,
\end{equation}
and their corresponding output parameters 
\begin{equation}
    \Y=\lbrace m_f, \vec{\chi}_f, \vec{v}_f \rbrace\,.
\end{equation}
The input spins are parameterized by their dimensionless spin magnitude $\chi_{i}=\|\chi_{i}\|$, spin tilt $\theta_{i}$, and azimuthal angle $\phi_{i}$ measured at $t=-100\,M$ before merger.
The final mass is scaled to the total mass at the simulation relaxation time, and the remnant dimensionless spin and recoil velocity are expressed in
Cartesian components. 
Input parameters \X lie within the boundaries of validity of \Remnant, namely $q\in\left[1,4\right]$, $\chi_{i}\leq0.8$ and
output parameters \Y are evaluated with \Remnant.

We utilize two ways of selecting \X points.
The first draws uniformly in
\begin{align}
    q&\in\left[1,4\right]\,,\label{eq:randomsamplingq}\\
\chi_{i}&\in\left[0,0.8\right]\,,\label{eq:randomsamplingchi}\\
\theta_{i}&\in\left[0,\pi\right]\,,\label{eq:randomsamplingthet}\\
\phi_{i}&\in\left[0,2\pi\right]\,.\label{eq:randomsamplingphi}
\end{align}
The second is the sparse grid method presented in Appendix A of Ref.~\cite{Blackman:2017dfb}, from which we construct $1528$ points, analogous to the $1528$ points used to train \Remnant~in the first place. 
The validation \Vset~and test \Sset~are populated with $10,000$ uniform draws.
The holdout set \Hset~is a combination of the $1528$ sparse-grid points
plus uniform draws up to a size of $100,000$ points.
The initial training set \Tset~is populated with the $1528$ sparse-grid points and will iteratively be augmented by points from the \Hset.

The training loss function is the mean absolute error (L1 norm) over all output parameters,
\begin{equation}
\Delta = \frac{1}{n}\|\Y^{\text{tr}} - \Y^{\text{pr}}\|\,,
\label{eq:MeanError}
\end{equation}
where $n$ is the number of output parameters, $\Y^{\text{tr}}$ is the vector of true values of the output parameters from \Remnant, and $\Y^{\text{pr}}$ are the predicted values. 
We monitor the loss over both training and validation sets during training and implement two conditions. 
Firstly, if the validation loss begins to monotonically increase for a period of $50$ training epochs, we stop training and return the weights to when the validation loss was at its minimum. 
This prevent overfitting features in the training data that are not also representative of the validation data, and therefore of the parameter space region as a whole. 
Second, we implement a learning rate scheduler which reduces the learning rate by a factor of $10$ whenever the training loss plateaus for a period of $50$ epochs. 
Larger learning rates allow the neural network optimizer to make bigger adjustments to the weights with each training step, while a smaller learning rate encourages fine-tuning when the learnable parameters are close to a local minimum.

\section{Hyperparameter and architecture optimization}
\label{sec:Hyperparameteroptimization}

\begin{table*}[t!]
    \centering
    \begin{tabular}{c||c|c|c|c|c|c|c}
    \hline
    \hline
    \multirow{2}{*}{Configuration} & Activation & Weight & \multirow{2}{*}{Normalization} & \multirow{2}{*}{Optimizer} & Initial Learning & Optimal Final & Final Loss \\
    & Function & Initialization & & & Rate & Loss & Standard Deviation\\
    \hline
    Mass & Softplus & HeNormal & No & Adam & $2.0\times10^{-4}$ & $1.5\times10^{-3}$ & $2.6\times10^{-1}$\\
    \hline
    Spins & Elu & glorot uniform & No & Nadam & $2.0\times10^{-4}$ & $1.1\times10^{-2}$ & $2.6\times10^{-1}$\\
    \hline
    Velocities & Elu & glorot uniform & No & Adam & $5.0\times10^{-4}$ & $4.0\times10^{-2}$ & $2.2\times10^{-1}$\\
    \hline
    6D & Softplus & HeNormal & No & Nadam & $5.0\times10^{-4}$& $1.9\times10^{-2}$ & $2.3\times10^{-1}$\\
    \hline
    7D & Softplus & glorot uniform & No & Nadam & $1.0\times10^{-3}$ & $1.8\times10^{-2}$ & $2.4\times10^{-1}$\\
    \hline
    \hline
    \end{tabular}
    \caption{Results for the functional hyperparameter optimization. We consider networks for different configurations of output parameters and list the optimal choice or value of each hyperparameter. We also list the optimal loss achieved as well as the standard deviation of final losses across the hyperparameter optimization stage.
    }
    \label{tab:FunctionalHyperparams}
\end{table*}

\begin{table*}[t!]
    \centering
    \begin{tabular}{c||c|c|c|c|c|c}
    \hline
    \hline
    Configuration & Hidden Layers & Neurons per Layer & Batch Size & Dropout & Optimal Final Loss & Final Loss Standard Deviation \\
    \hline
    Mass & $3$ & $10$ & $64$ & No & $2.0\times10^{-4}$ & $3.0\times10^{-2}$ \\
    \hline
    Spins & $5$ & $200$ & $256$ & No & $1.1\times10^{-2}$ & $9.7\times10^{-2}$ \\
    \hline
    Velocities & $8$ & $50$ & $128$ & No & $1.8\times10^{-2}$ & $1.2\times10^{-1}$ \\
    \hline
    6D & $5$ & $100$ & $128$ & No & $1.8\times10^{-2}$ & $1.3\times10^{-1}$ \\
    \hline
    7D & $5$ & $100$ & $128$ & No & $1.7\times10^{-2}$ & $1.2\times10^{-1}$ \\
    \hline
    \hline
    \end{tabular}
    \caption{Same as Table~\ref{tab:FunctionalHyperparams} but for the network architecture and specifically the size and shape parameters.}
    \label{tab:SizeHyperparams}
\end{table*}

In this section, we introduce a systematic approach to optimize the neural network architecture and hyperparameters through the hyperparameter optimization package \optuna \cite{optuna_2019}, which uses a Tree-of-Parzen-Estimator \cite{bergstra2011algorithms} algorithm to update hyperparameter choices based on estimates of the final loss function evaluated on the unseen \Sset.
We split the network optimization problem into two separate steps, first for the hyperparameters and second for the architecture (size and shape). 
We carry out the optimization separately for different configurations of output parameters: remnant mass (1D), remnant spin vector (3D), kick velocity (3D), remnant spin and kick velocity (6D), and all together (7D).

The optimization procedure of \optuna proceeds as follows. 
The algorithm draws initial samples of hyperparameter combinations from the prior of options given by the user, and trains these models. 
Based on the final loss values, it splits these samples into one ``good'' group of hyperparameter combinations and one ``bad'' group, and then uses kernel density estimators to model the probability distributions of these two groups.
It then iteratively proposes new samples that maximize the ratio of the ``good'' probability distribution to the ``bad'' one, evaluating these new models and updating the groups. The \optuna algorithm further relies on a number of hyperparameters, including: the prior weights for hyperparameters; the number of startup trials which are sampled randomly from the prior; the number of trials run; and the number of samples used to calculate the expected improvement, which is then maximized to choose the next samples. Configuration files for our analyses listing all choices are available at \cite{gwbonsai}.

We begin with the functional hyperparameters as their impact on the fit is more complex, while fixing the size and shape parameters. 
We fix the number of hidden layers to $4$ with no dropout layers, neurons per layer to $100$, and batch size to $128$ (more details about the role of these parameters are provided later).
We also use a fixed training set size of 10,000 points.
We study the effect of the functional hyperparameters as follows. Note that these functional hyperparameters are optimized over simultaneously in this step.
\begin{enumerate}[(i)]
    \item The \emph{activation function} is applied to data passing across each neuron in the hidden layers, and adds non-linearity to the network which allows it to universally approximate functions. Different activation functions and their gradients (which need to be stable and easy to compute) will work better for different problems, and we consider: \texttt{relu}, \texttt{sigmoid}, \texttt{softplus}, \texttt{softsign}, \texttt{tanh}, \texttt{selu}, \texttt{elu}, \texttt{exponential}, and \texttt{LeakyReLU}. 
    \item \emph{Weight initialization} assigns initial values to the weights of neurons based on a distribution, and acts as a seed for training. 
    A suitable choice of weight initialization can allow for faster and more stable training by addressing vanishing or exploding gradients, and we consider a choice of \texttt{Glorot uniform}, \texttt{HeNormal} and \texttt{HeUniform} distributions. 
    \item We consider whether to add a \emph{normalization layer} within the network, which scales the inputs to zero mean and unit variance. Though we scale the input and output training data in the same way, renormalization in the middle of the network can help stabilize the gradients of the activation function and allow for smoother training. 
    \item The \emph{optimizer} is the algorithm that updates the neuron weights during training; we explore: \texttt{Adam}, \texttt{Adamax}, \texttt{Nadam}, and \texttt{Ftrl}. 
    \item While we use an adaptive learning rate as described in Sec.~\ref{sec:ModelStructureandDatasets}, we consider the \emph{initial learning rate} to adjust how aggressive the initial training adjustments are, allowing a discrete set of values: $[1\times10^{-5},~2\times10^{-5},~5\times10^{-5},~1\times10^{-4},~2\times10^{-4},~5\times10^{-4},~1\times10^{-3}]$.
\end{enumerate}

Results for the hyperparameter optimization are shown in Table~\ref{tab:FunctionalHyperparams}.
For each configuration of output parameters, we list the optimal choice of hyperparameter, the resulting optimal final loss, and the standard deviation of losses across the functional hyperparameter optimization stage. The final loss for the optimal configuration is similar for all configurations other than the mass-only network, for which it is a factor of 10 lower, likely because it is the simplest one.
For all configurations, the optimal loss is around an order of magnitude lower than the standard deviation, indicating that the optimization process indeed found a well-performing minimum setting.

With the optimal functional hyperparameters from Table~\ref{tab:FunctionalHyperparams}, we turn to the network architecture and specifically the size and shape parameters of the network layers. For this relatively simple regression problem, we restrict to a fully-connected multi-layer perceptron with all layers the same size.
\begin{enumerate}[(i)]
\item  Qualitatively speaking, the \emph{number of neurons per layer} corresponds to the number of ``features" we expect in the data, and the \emph{number of hidden layers} to how complex those features are. 
Optimizing over these parameters is crucial for avoiding over- and under-fitting. 
We consider layer sizes between $10$ and $200$ neurons and the number of hidden layers between $2$ and $12$ (excluding input, output, normalization, and dropout layers). 
\item  The \emph{batch size}, which is the number of points passed to the network simultaneously during training, impacts the tendency to over-fit if too small.
We consider values in powers of $2$ between $64$ and $512$. 
\item  We explore the impact of a \emph{dropout layer}, which probabilistically discards neuron information during training. 
Dropout layers can help avoid over-fitting, so we optimize over whether to include one, and what dropout probability it has, with a uniform prior between 0 and 1.
\end{enumerate}
Results for the network architecture optimization step for each configuration of output parameters is shown in Table~\ref{tab:SizeHyperparams}. 
The optimal final losses and standard deviations are again relatively similar for all configurations except the mass-only one. 
The optimal losses are still around an order of magnitude lower than the standard deviations. 

The optimization results of Tabs.~\ref{tab:FunctionalHyperparams} and~\ref{tab:SizeHyperparams} elucidate the role of the various settings as well as the structure of the problem.
Many results are similar for the different output configurations, indicating that they are the generally preferred settings.
For example, none of the configurations benefit from a normalization or dropout layer,
and all configurations are best trained with an \texttt{Adam} optimizer or its closely related variant \texttt{Nadam}, which uses the \texttt{Adam} algorithm with Nesterov momentum to correct for overly large gradient jumps. 
The preferred activation functions are \texttt{softplus}, which is a smooth approximation to the often-used \texttt{relu}, and \texttt{elu}, which is also similar to the \texttt{relu} activation but allowing for negative values of neuron output which can speed up learning. 
The optimal weight initialization distribution is a mixture of the commonly used \texttt{glorot uniform} and \texttt{HeNormal}, which is suited to \texttt{relu} and related activation functions \cite{pmlr-v9-glorot10a}.

The optimal solutions given in Table~\ref{tab:FunctionalHyperparams} are relatively stable to the choice of activation function, as throughout sampling the optimal activation function generally led to lower final losses even as other functional hyperparameters varied. 
The performance of different weight initialization and optimizer settings were more dependent on other hyperparameters, and the \texttt{Adam} and \texttt{Nadam} optimizers performed similarly for all configurations. 
We also found a positive correlation between the batch size and number of neurons per layer. 
This is expected, as larger layers will need more information to update their weights and train efficiently.

\begin{figure*}[]
    \centering
    \includegraphics[width=0.8\textwidth]{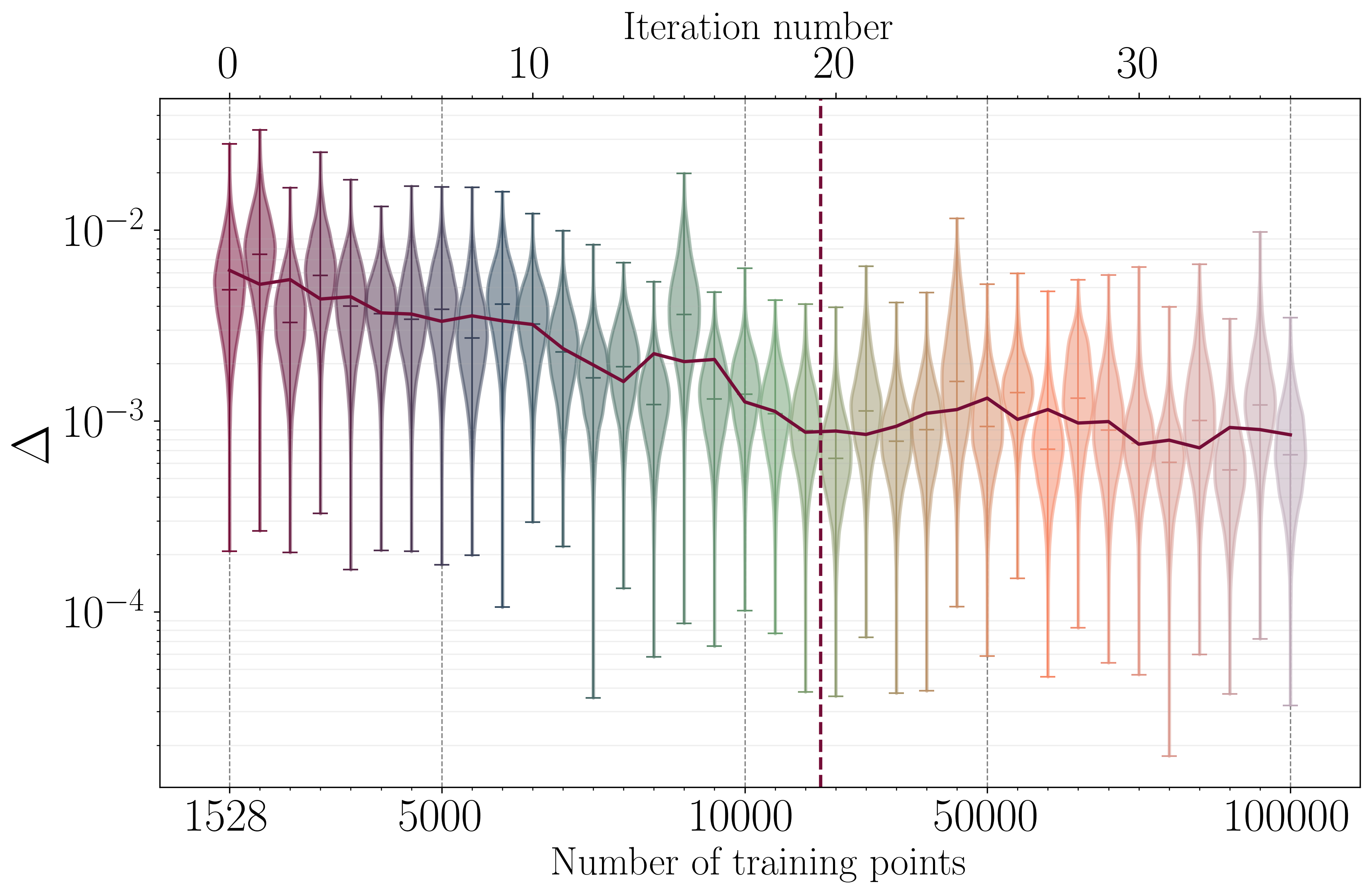}
    \caption{Distribution of the mean error $\Delta$ over the unseen \Sset~(violins) as a function of the training dataset size for the 7D configuration network and the training procedure detailed in Sec.~\ref{sec:TrainingDatasetSizeOptimization}. As the number of training points increases, the mean error tends to decrease and plateaus after around $19$ iterations (dashed line). In purple solid, we show the running mean over three iterations of the median error, which we use to assess where the plateau begins. We select 20,000 points (roughly the size of the 19th iteration) as the optimal training dataset size.}
    \label{fig:7DMeanErrorViolin}
\end{figure*}

Differences between the optimal hyperparameters for the different configurations further reveal information about the properties of the parameter space.
The remnant spin configuration requires the most neurons per layer and the largest batch size, suggesting a feature-rich parameter space.
The kick velocity optimizes to the largest number of hidden layers, suggesting more complex features to fit. 
However, when spin and velocity are combined together in a single 6D configuration, or as part of the full 7D network, this complexity is reduced, suggesting that the different output parameters are related to each other. 
The mass-only configuration is the least feature-rich, requiring the fewest and smallest layers, and the smallest batch size.
This result is expected given that the final mass is always close to $1$, further highlighting the sensibility of the optimization framework.
Finally,
all configurations except the full 7D one are optimized with a medium-valued initial learning rate of $2-5\times10^{-4}$, while the full 7D configuration benefits from a larger initial learning rate of $1\times10^{-3}$ for bigger weight adjustments. 

\section{Training Dataset Size Optimization}
\label{sec:TrainingDatasetSizeOptimization}

\begin{figure*}[]
    \centering
    \includegraphics[width=0.49\textwidth]{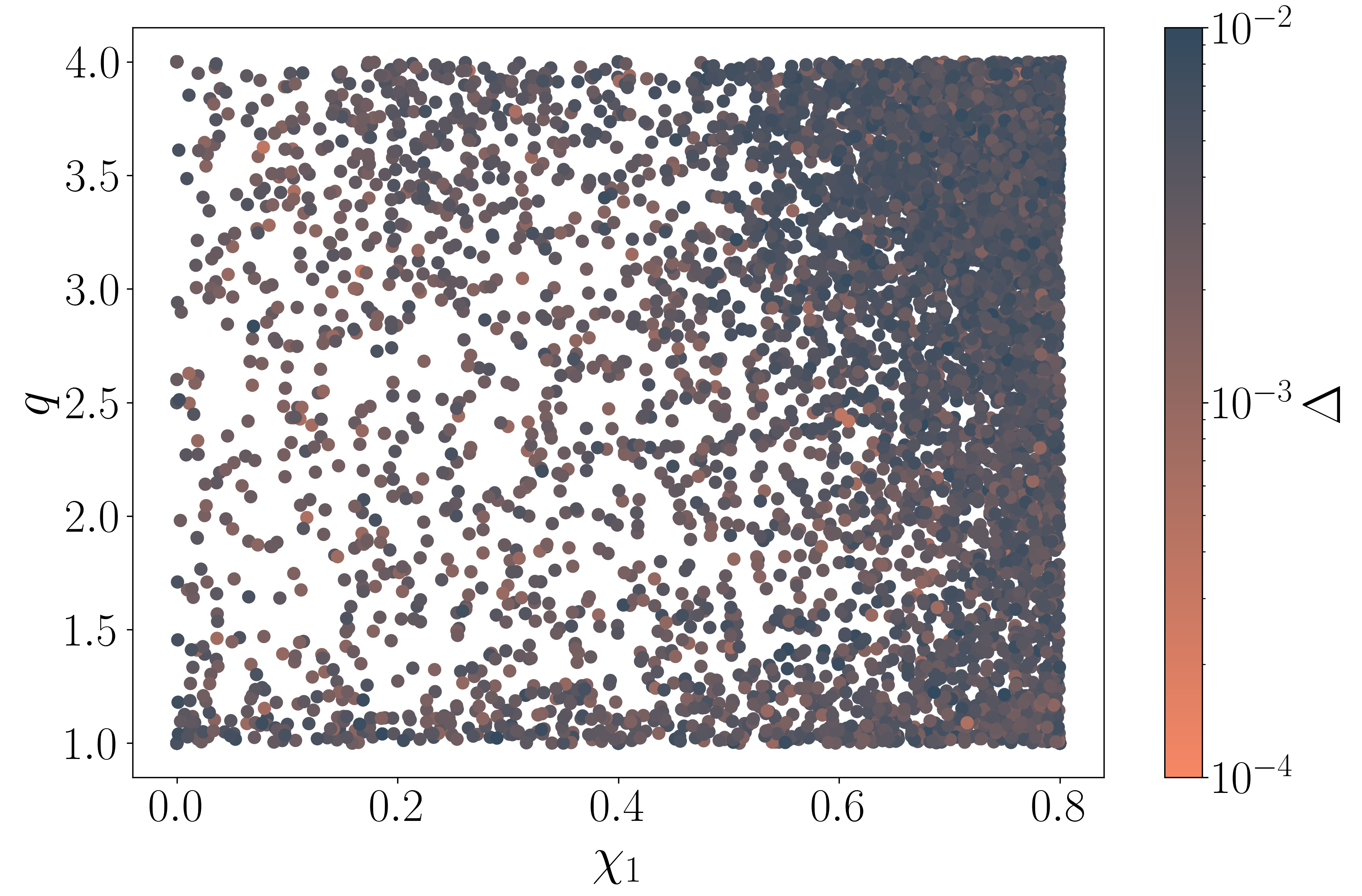}
    \includegraphics[width=0.49\textwidth]{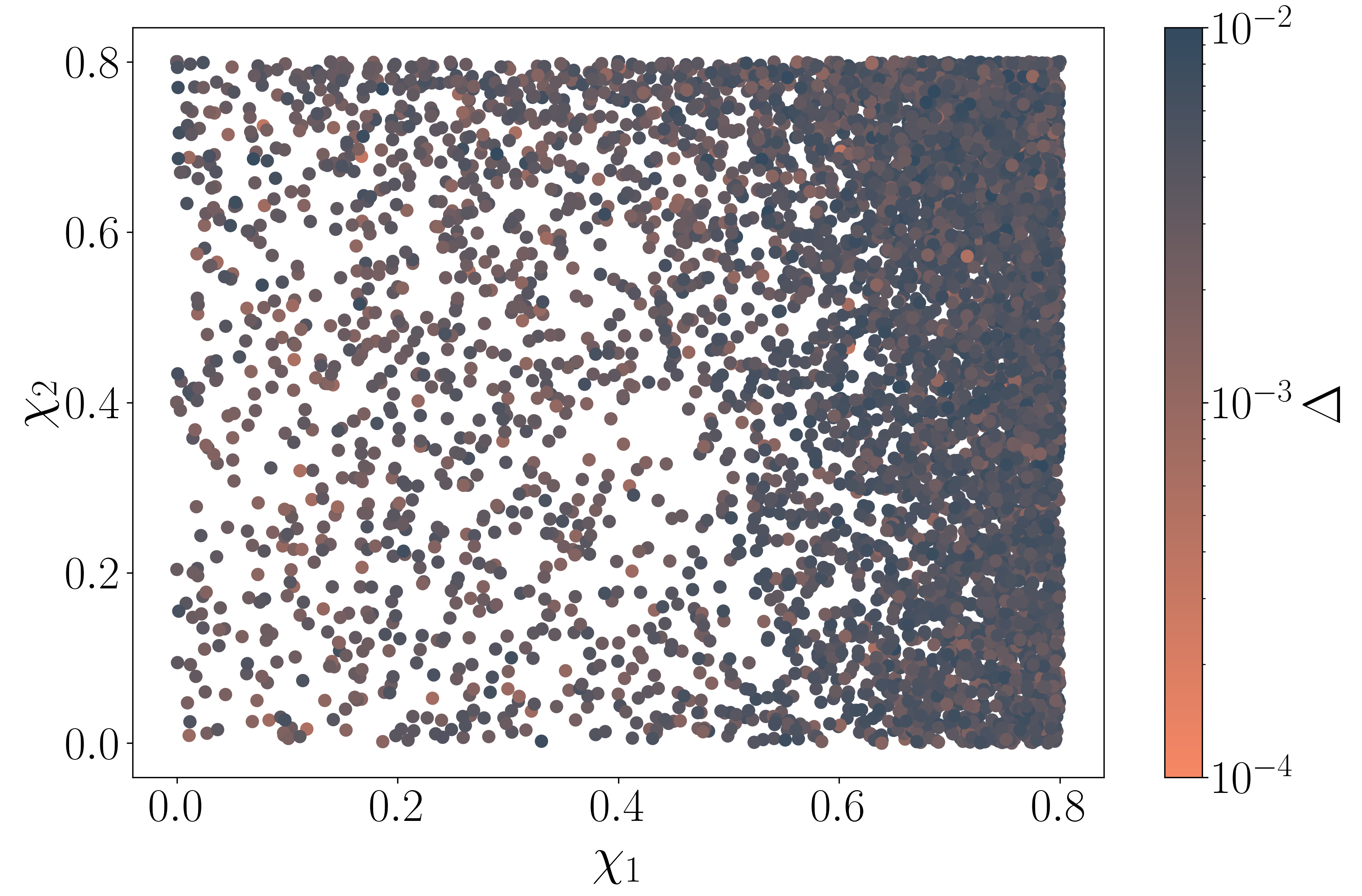}\\
    \includegraphics[width=0.49\textwidth]{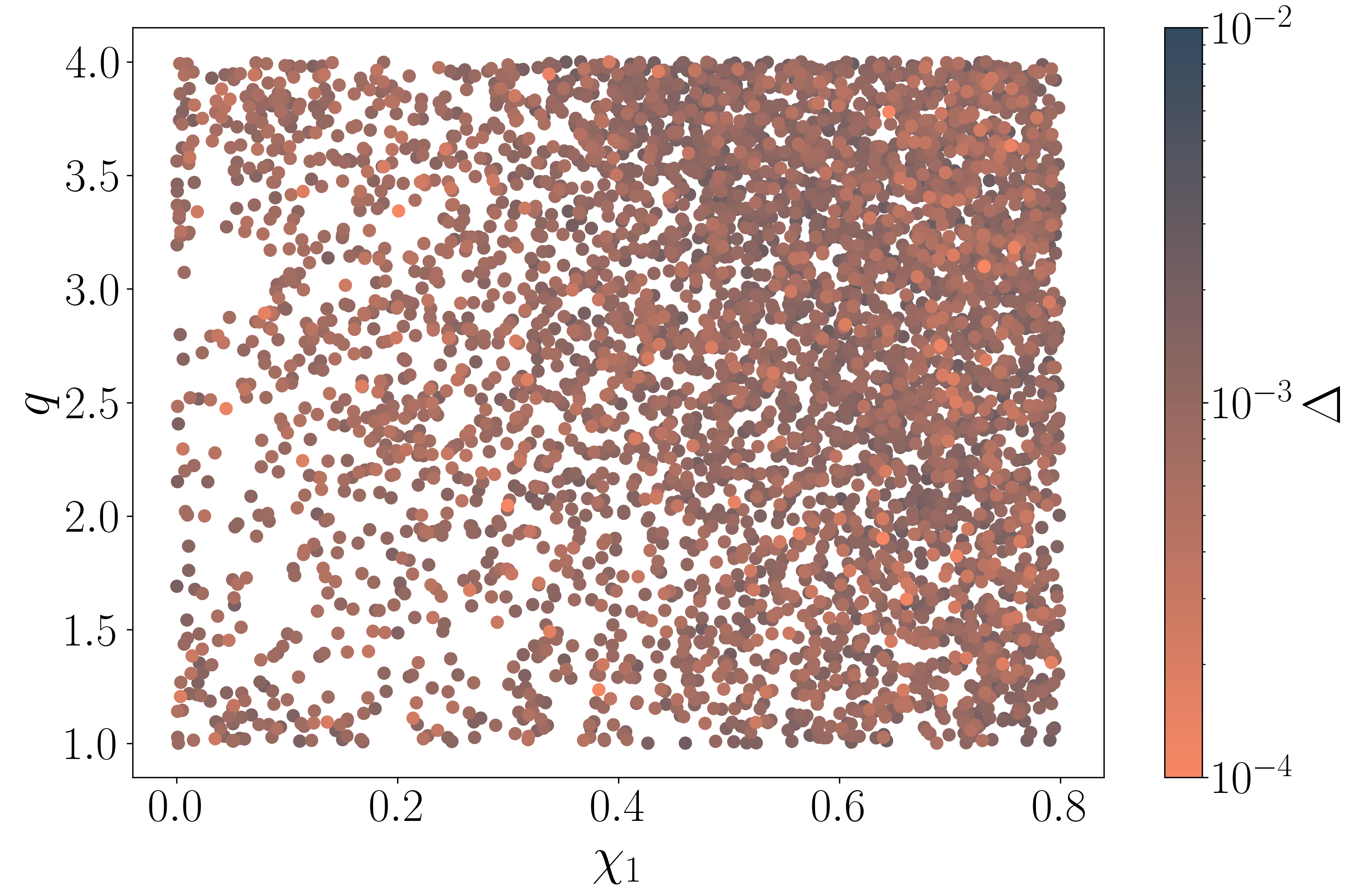}
    \includegraphics[width=0.49\textwidth]{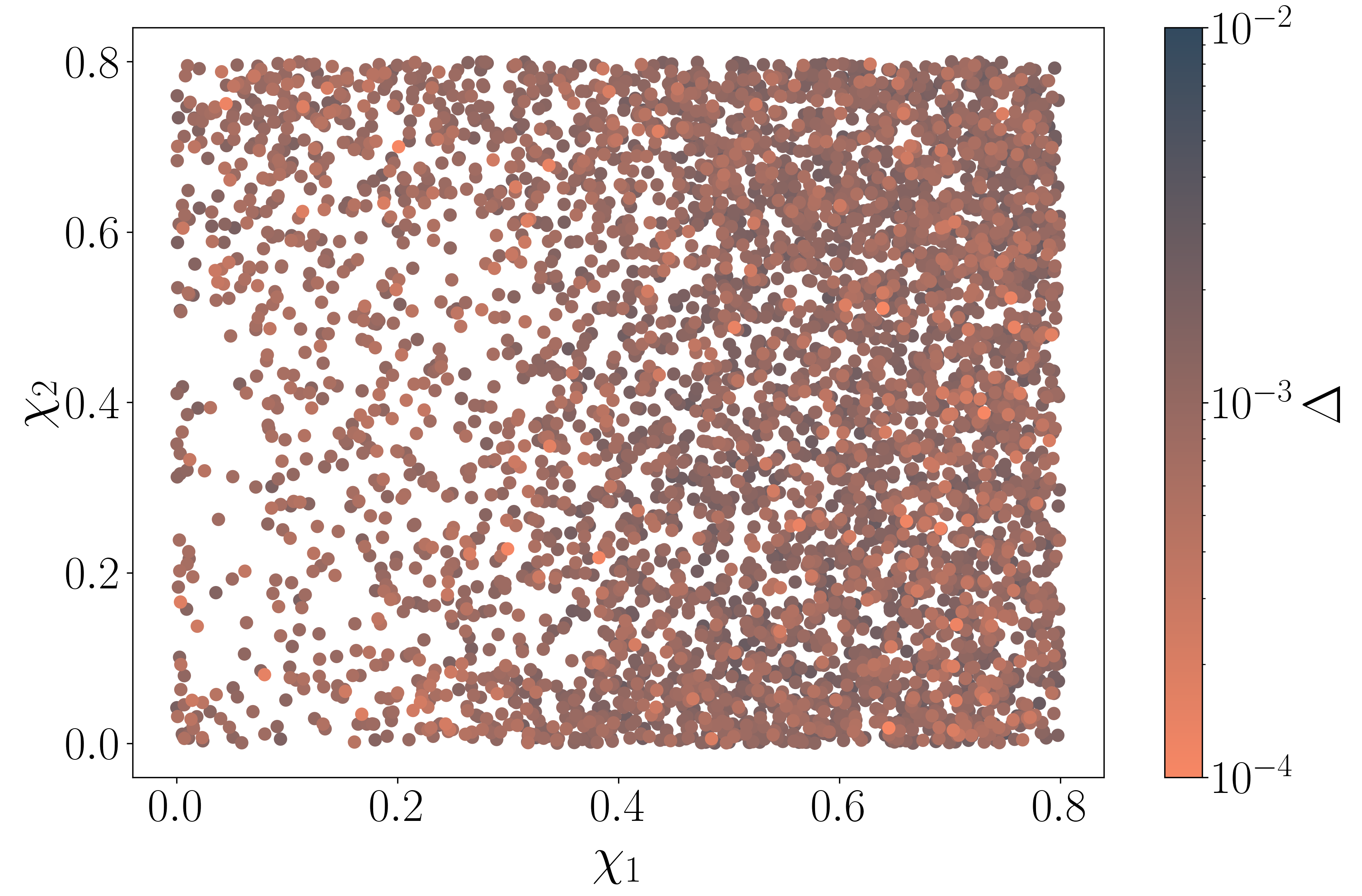}
    \caption{Distribution of the $1,000$ worst-performing points in the \Tset~across the parameters in mass ratio-primary spin magnitude (left column) and spin magnitudes (right column) for the full 7D configuration after $10$ (top row) and $19$ (bottom row) iterations.
    Each point is colored by its mean error $\Delta$. 
    These points are not part of the training set for their respective iteration, but are then added to the training set for the next iteration.
    The 10th iteration worst-performing points are concentrated in the high mass ratio and large spin magnitude areas. 
    By the 19th iteration (where the performance plateau occurs), the worst-performing points are distributed across a wider range of spin magnitudes in the training region, more evenly in mass ratios,  and with lower mean errors.
    This suggests that the iterative enrichment of the training set effectively samples the parameter space by dynamically selecting the most representative points.
    }
    \label{fig:IterativeParamSpace}
\end{figure*}

\begin{table}[t!]
    \centering
    \begin{tabular}{c||c|c|c}
    \hline
    \hline
    Configuration & Training Points & Epochs & Final Loss \\
    \hline
    Mass & $15,000$ & $5,100$ & $6.4\times10^{-5}$\\
    \hline
    Spins & $30,000$ & $4,320$ & $5.3\times10^{-5}$\\
    \hline
    Velocities & $45,000$ & $6,086$ & $4.6\times10^{-5}$\\
    \hline
    6D & $30,000$ & $5,883$ & $5.0\times10^{-5}$ \\
    \hline
    7D & $20,000$ & $6,003$ & $4.5\times10^{-5}$ \\
    \hline
    \hline
    \end{tabular}
    \caption{Settings and performance of the final networks for each configuration of output parameters. Besides the optimal hyperparameters and architecture listed in Tables~\ref{tab:FunctionalHyperparams} and~\ref{tab:SizeHyperparams} respectively, here we list the optimal training size, determined in Sec.~\ref{sec:TrainingDatasetSizeOptimization}, the training epochs, and the final loss achieved.}
    \label{tab:FinalHyperparams}
\end{table}

Having selected values for the network size and shape as well as the functional hyperparameters, we now assess how much training data is required. 
We follow an iterative approach:
\begin{enumerate}
\item Begin with a training \Tset~containing the $1528$ sparse-grid points.
\item Train a neural network for $2,000$ epochs, chosen to balance computational cost with obtaining representative results.
\item Evaluate the fit accuracy on the holdout \Hset~and test \Sset~sets.
\item Choose the worst performing points from the \Hset~and add them to the \Tset~(up to next multiple of $500$, or next multiple of $5,000$ if more than $10,000$ training points).
\item Retrain the neural network.
\item Repeat steps 2-5 until \Tset=\Hset.
\end{enumerate}

Figure~\ref{fig:7DMeanErrorViolin} shows the distribution of mean error $\Delta$, Eq.~\eqref{eq:MeanError}, for the \Sset~for the 7D network configuration; we obtain qualitatively similar trends for other configurations. 
The mean error overall improves with increasing size of the training set, as the violins trend downwards with each iteration. 
The running mean of the median $\Delta$ over 3 iterations plateaus after ${\sim}19$ iterations, indicating diminishing returns from further increasing the amount of training data. 
We, therefore, identify $20,000$ points as the optimal training size for this configuration.

We repeat this process for all network configurations and identify a training dataset size that balances accuracy with training cost. 
Results are shown in Table~\ref{tab:FinalHyperparams}. 
The mass-only configuration reaches the loss plateau with the smallest amount of training data.
This is consistent with the findings of Sec.~\ref{sec:Hyperparameteroptimization} that the final mass is the simplest to train, requiring the lightest network settings. 
On the other side, the kick velocity requires the most training data, to represent the complexity of the parameter space features, again consistent with Sec.~\ref{sec:Hyperparameteroptimization}. 
Interestingly, again the full 7D configuration network requires fewer points (lighter settings) than the technically lower-dimensional networks for the spins and/or velocities.
This again suggests that including all parameters in a single network adds information that simplifies the problem.
Even in this case, however, the training size required, ${\cal{O}}(10^4)$, is about an order of magnitude higher than the size of current NR catalogs and the training size of the original surrogate \Remnant, ${\cal{O}}(10^3)$.

Figure~\ref{fig:7DMeanErrorViolin} and Table~\ref{tab:FinalHyperparams} quote the mean error over all output parameters for a given configuration.
We further examine the mean error for subsets of the output parameters of each configuration, for example, the final mass, remnant spin magnitude or angle, and kick velocity magnitude or angle, as appropriate. 
Interestingly, we find that the training dataset size at which the loss plateaus is independent of which output parameters we consider. 
This suggests that there is an optimal training dataset for all output parameters and which improves the network accuracy as a whole.

The iterative addition of training data further allows us to explore which regions of the parameter space contribute to the loss of accuracy.
Figure~\ref{fig:IterativeParamSpace} shows the points that get added to the training set.
The top row shows the mean error $\Delta$ for the $1,000$ worst-performing points across the \Tset~after $10$ iterations. 
The worst-performing points are concentrated in areas with high mass ratios and spin magnitudes. 
These $1,000$ points are then added to the training dataset for the next iteration. 
In the bottom row, we show the same mean error but for the $1,000$ worst-performing points in the \Tset~after $19$ iterations (which is where the plateau in improvement in Fig.~\ref{fig:7DMeanErrorViolin} occurs).  
The worst-performing points are more evenly spread throughout the parameter space and the mean errors are lower overall.
This suggests that the iterative procedure for adding training data identifies problematic parameter space regions and efficiently fills them up.
By the time the performance plateau is reached, errors are not only reduced but also more evenly distributed across parameters.

\section{Final Model Evaluation}
\label{sec:FinalModelEvaluation}

Having optimized the network architecture, hyperparameters, and training dataset size, we now train the final models for each configuration of output parameters.
Specifically, we use the hyperparameters and size and shape parameters listed in Tables~\ref{tab:FunctionalHyperparams} and~\ref{tab:SizeHyperparams} respectively and the training dataset size in Table~\ref{tab:FinalHyperparams}.
We use an adaptive learning rate and the early stopping condition described in Sec.~\ref{sec:ModelStructureandDatasets}; the resulting final number of training epochs is in Table~\ref{tab:FinalHyperparams}.
Also listed in Table~\ref{tab:FinalHyperparams} is the final loss achieved after each network is fully trained. 
We obtain comparable losses for each configuration, with the 7D configuration performing marginally better.
Coupled to the simpler settings in Tables~\ref{tab:FunctionalHyperparams} and~\ref{tab:SizeHyperparams}, we therefore conclude that the 7D configuration is preferred and present results only for this network in what follows, which we label as \RemnantNN.

\begin{figure*}[]
    \centering
    \includegraphics[width=0.49\textwidth]{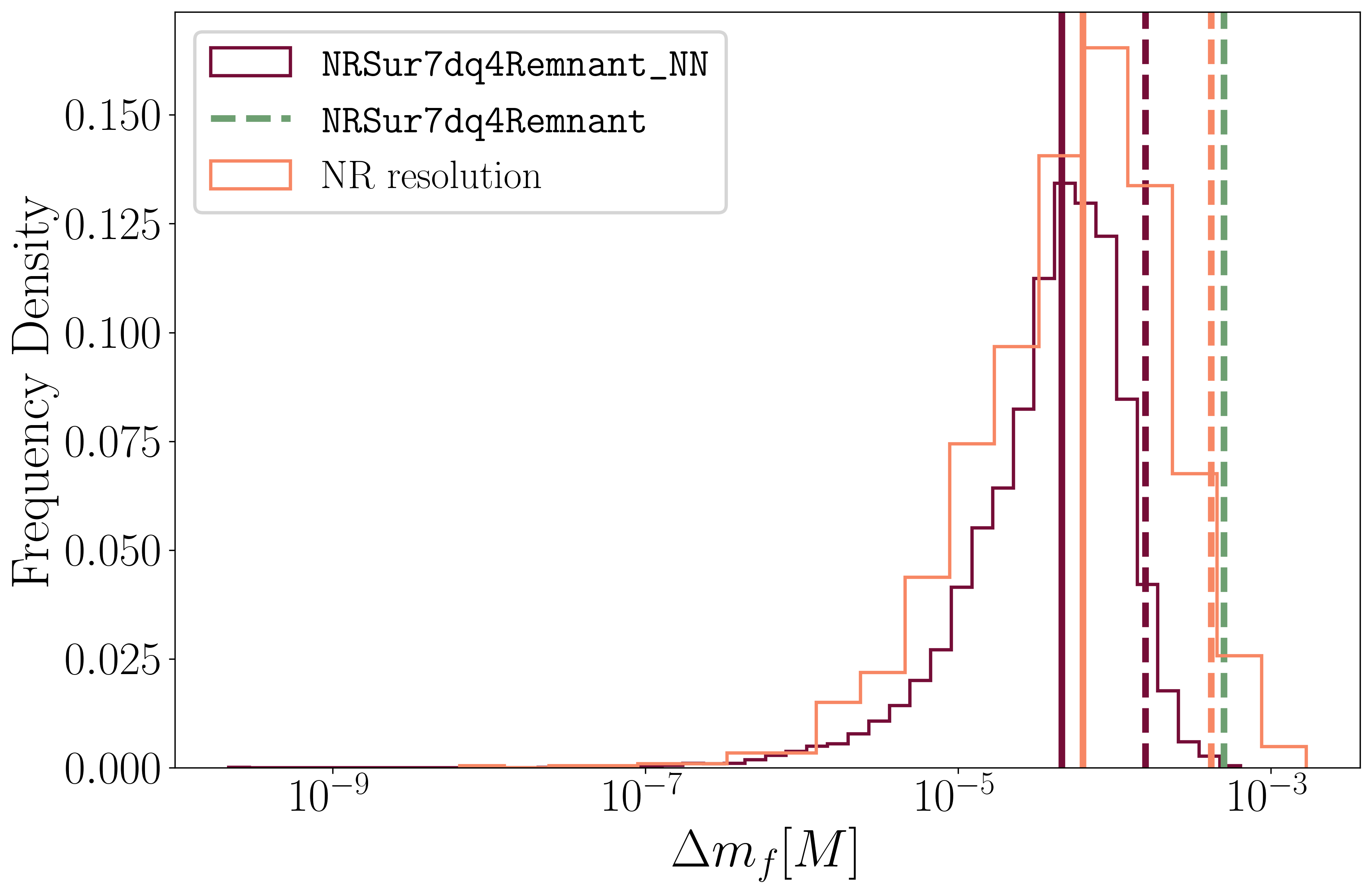}
    \includegraphics[width=0.49\textwidth]{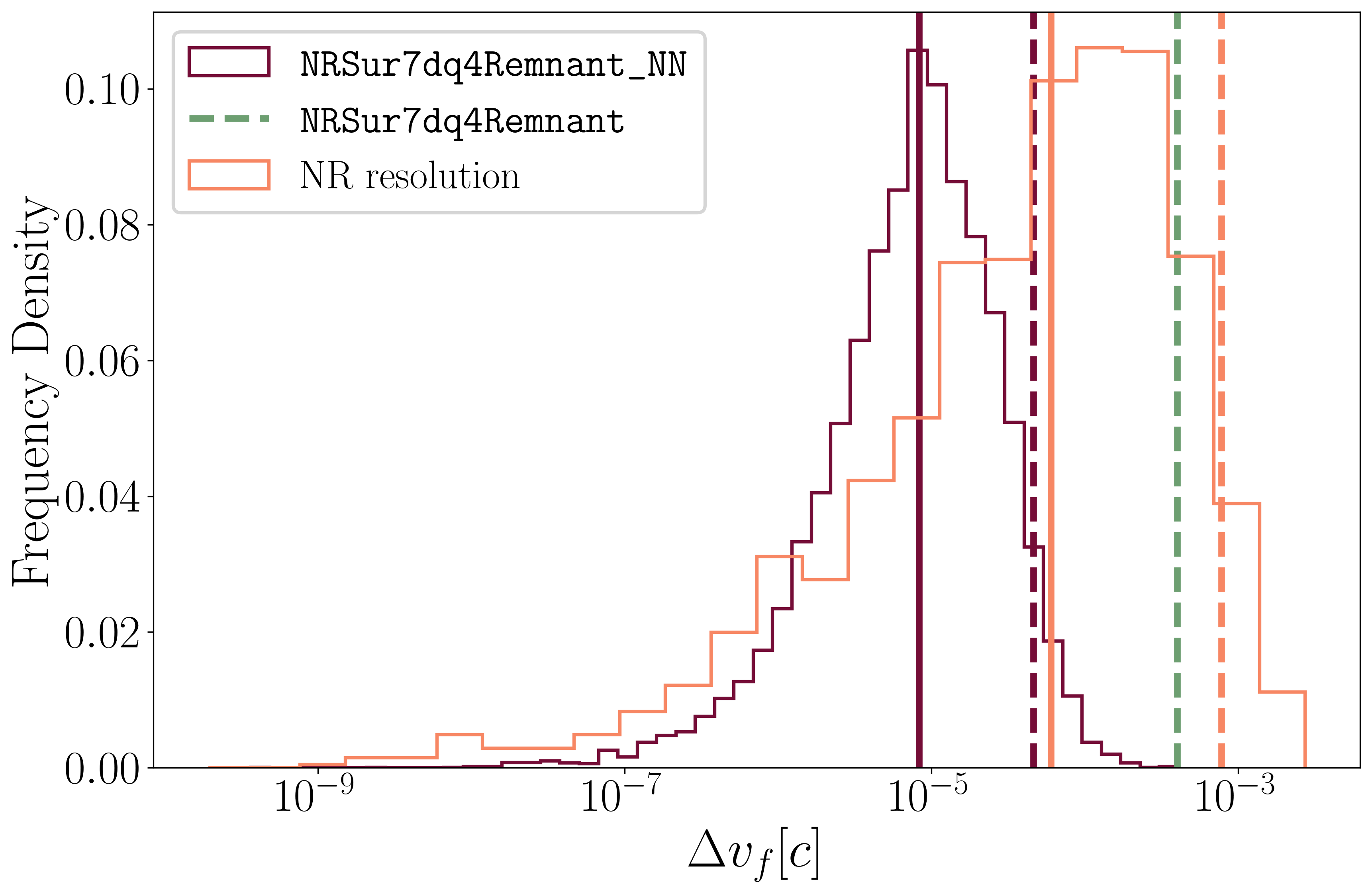}\\
    \includegraphics[width=0.49\textwidth]{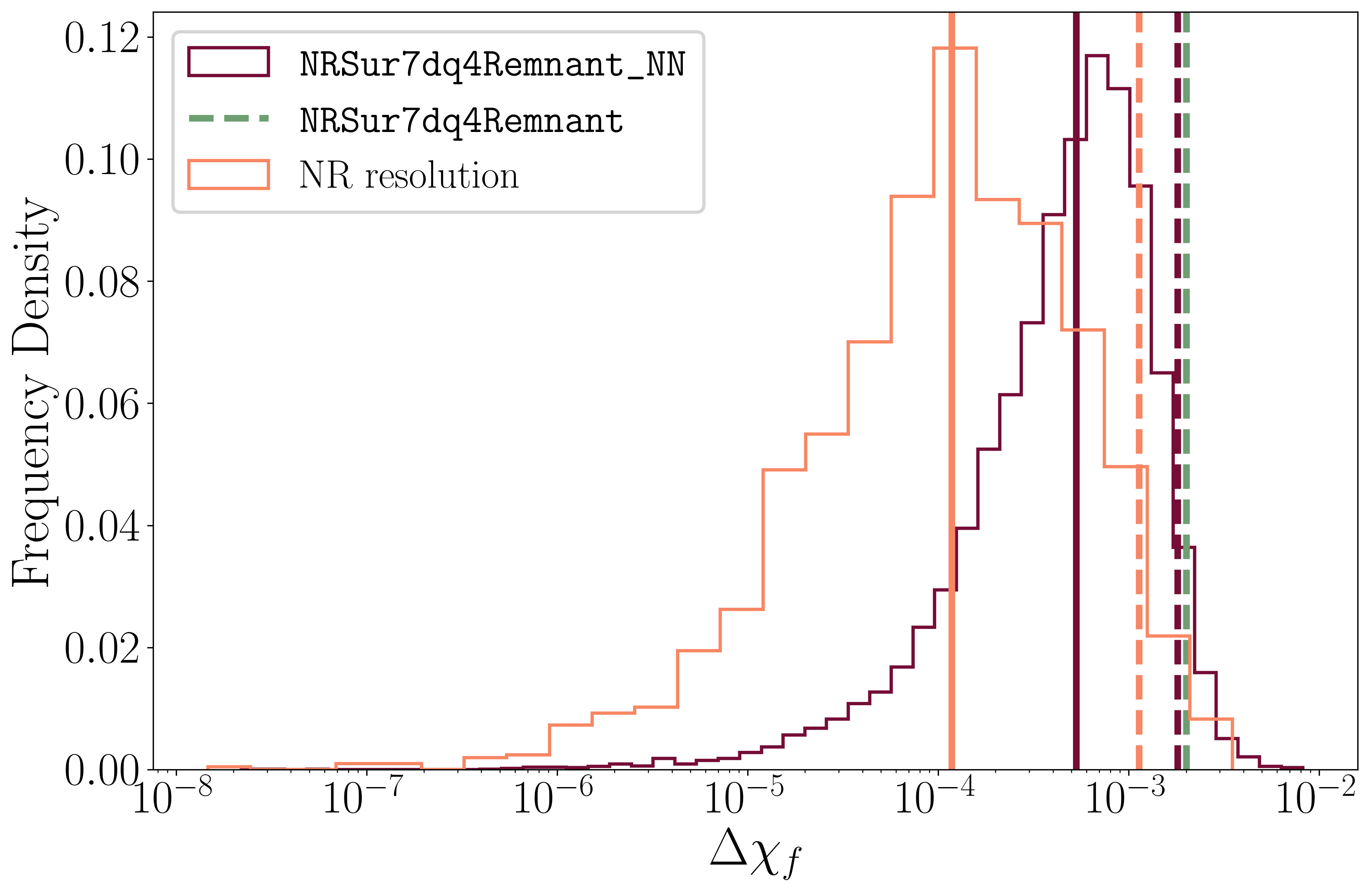}
    \includegraphics[width=0.49\textwidth]{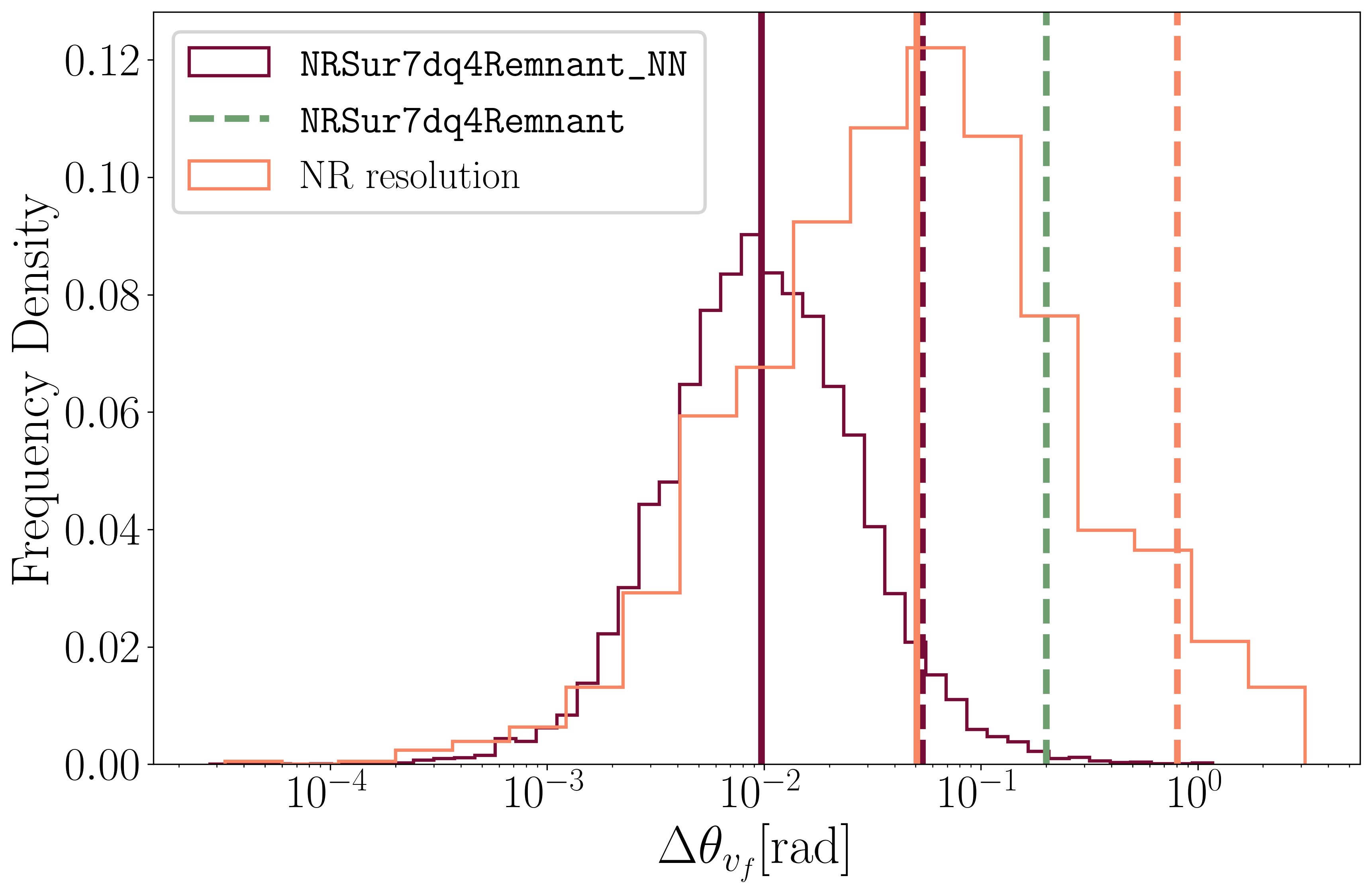}\\
    \includegraphics[width=0.49\textwidth]{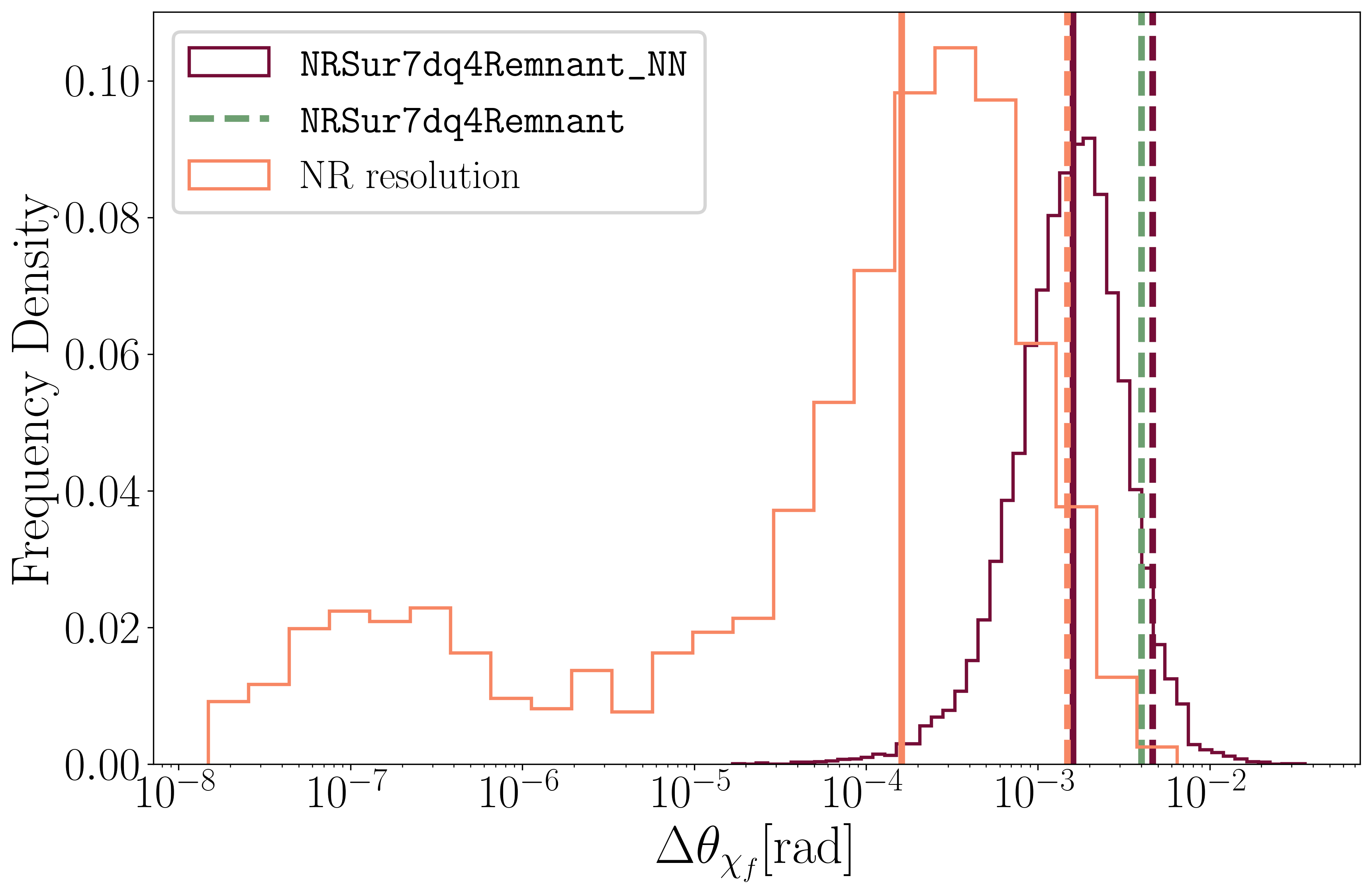}
    \includegraphics[width=0.49\textwidth]{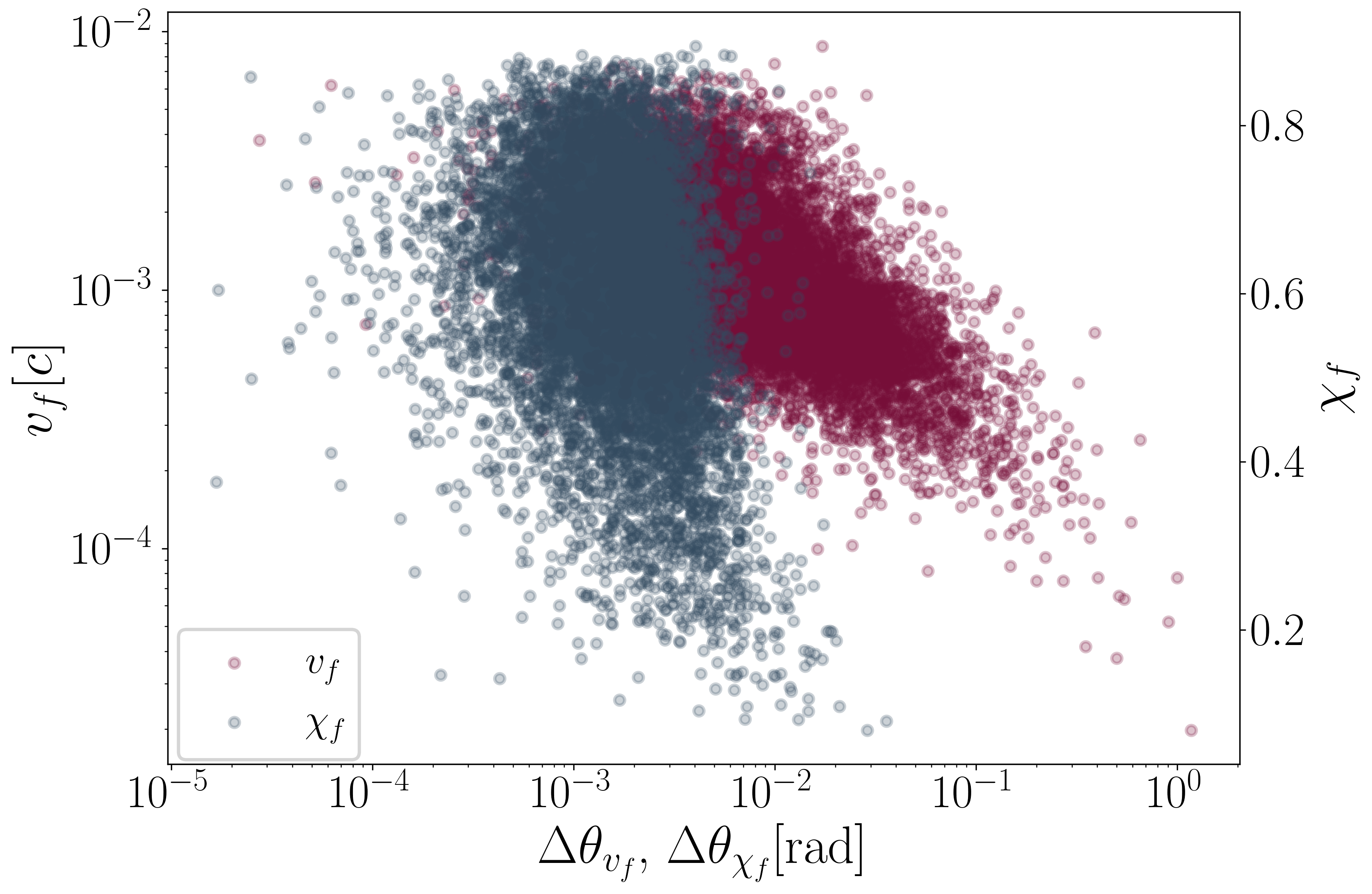}
    



    \caption{Errors over the \Sset~for the remnant mass, $\Delta m_f$, kick magnitude, $\Delta v_f$, spin magnitude, $\Delta \chi_f$, kick direction $\Delta \theta_{v_f}$, and spin direction, $\Delta \theta_{\chi_f}$. 
    Besides error histograms, in the bottom right panel we also show the joint distribution of magnitude and direction errors for both the kick velocity (purple) and spin vectors (blue). 
    Results histograms for \RemnantNN~compared to the \Remnant~surrogate it was trained on are shown in purple, with medians (solid) and 95th percentiles (dashed) shown as purple vertical lines. Green dashed vertical lines indicate the 95th percentile errors of \Remnant~with respect to the NR it is built upon; these errors are sometimes larger than the reported \RemnantNN~errors as the latter is compared with \Remnant~instead of NR. We also show the distributions of NR resolution errors in orange, with medians (solid) and 95th percentiles (dashed) shown as vertical orange lines.
    }
    \label{fig:FinalAccuracyResults}
\end{figure*}

The accuracy of \RemnantNN~is assessed in Fig.~\ref{fig:FinalAccuracyResults}.
We plot the distribution of errors across the unseen \Sset~for output parameters defined as
\begin{align}
\Delta m_f &= \left|m_f^{\text{tr}}-m_f^{\text{pr}}\right|\,,\\
\Delta x &= \left|\vec{x}^{\text{tr}}-\vec{x}^{\text{pr}}\right|\,,\\
\Delta \theta_{{x}} &= \cos^{-1}\left(\hat{{x}}^{\text{tr}}\cdot\hat{{x}}^{\text{pr}}\right)\,,
\end{align}
where $\vec{x}$ represents either of the vectors $\vec{\chi}_f$ or $\vec{v}_f$, and true and predicted values refer to \Remnant~and \RemnantNN~respectively. 
We also plot the medians and 95th percentiles of these distributions as solid and dashed purple vertical lines, respectively, and the 95th percentiles of the errors of \Remnant~ with respect to the underlying NR data it is based on as dashed green vertical lines. 
We list these 95th error percentiles in Table~\ref{tab:95thPercentiles}.
For reference, we also plot the resolution error of the NR simulations that \Remnant~was based on (computed from the two highest NR resolutions available. Due to different frame choices when computing the resolution errors, our resolution errors differ slightly from Ref.~\cite{Varma:2019csw}. Here, we use the publicly-available metadata \cite{Boyle:2019kee} to compute errors in the initial frame of the simulations, whereas in Ref.~\cite{Varma:2019csw} the resolutions are rotated into a frame which aligns with the coorbital frame $100M$ before merger at the time where the input masses and spins are measured.
Similar to \Remnant~in Ref.~\cite{Varma:2019csw}, the NR resolution errors are comparable or slightly higher than the surrogate model errors for the remnant mass and kick velocity magnitude and direction, suggesting that the surrogate building is not a dominant source of error. 
For the remnant spin quantities, the resolution errors are lower than those of our model. 
The distribution of errors over the \Sset~is similar to errors calculated over the larger \Tset, suggesting that the \Sset is sufficiently dense to quantify the errors over the 7D space.

Compared to \Remnant~that it was trained on, \RemnantNN~is extremely accurate. 
For the final mass and recoil velocity, the 95th percentiles of \RemnantNN~are less than those of \Remnant. 
For the kick velocity magnitude specifically, the 95th percentile error is almost an order of magnitude less than \Remnant. 
We also find improvements of roughly a factor of $3$ and $2$ in the remnant mass error and kick velocity direction, respectively.
For the final spin, the errors are comparable.
This suggests that \RemnantNN~incurs only minimal additional errors on top of those of \Remnant.

\begin{table*}[]
    \centering
    \begin{tabular}{c||c|c|c}
    \hline
    \hline
   Error & \RemnantNN~ & \Remnant~ & NR Resolution Error \\
     Quantity & 95th Percentile & 95th Percentile & 95th Percentile\\
    \hline
    $\Delta m_f [M]$ & $1.6\times10^{-4}$ & $5\times10^{-4}$ & $4.1\times10^{-4}$ \\
    \hline
    $\Delta \chi_f$ & $1.8\times10^{-3}$ & $2\times10^{-3}$ & $1.1\times10^{-3}$\\
    \hline
    $\Delta \theta_{\chi_f} [\text{rad}]$ & $4.6\times10^{-3}$ & $4\times10^{-3}$ & $1.5\times10^{-3}$\\
    \hline
    $\Delta v_f [c]$ & $4.6\times10^{-5}$ & $4\times10^{-4}$ & $7.8\times10^{-4}$ \\
    \hline
    $\Delta \theta_{v_f} [\text{rad}]$ & $5.4\times10^{-2}$ & $2\times10^{-1}$ & $8.0\times10^{-1}$ \\
    \hline
    \hline
    \end{tabular}
    \caption{95th percentile errors as shown by the dashed vertical lines in Fig.~\ref{fig:FinalAccuracyResults}. The first column shows the errors of \RemnantNN~with respect to \Remnant,  the second column shows the errors of \Remnant~with respect to the underlying NR simulations, and the final column shows the NR resolution errors calculated between the two highest resolution levels.} 
    \label{tab:95thPercentiles}
\end{table*}

In the bottom right panel of Fig.~\ref{fig:FinalAccuracyResults}, we show a scatter plot of the remnant spin magnitude $\chi_f$ against the spin direction error $\Delta\theta_{\chi_f}$ (left y-axis), and the kick velocity magnitude $v_f$ against the kick direction error $\Delta\theta_{v_f}$ (right $y$-axis). 
Reference~\cite{Varma:2019csw} reported a larger angular error for larger kick velocity magnitudes; we confirm this result and extend it to the spin vector.
This is likely because larger magnitude vectors are easier to fit, as our loss function is the mean error over the Cartesian vector coordinates. 
Therefore, a larger error in angle for a smaller magnitude vector will have less of an impact on the overall loss.

\begin{figure}[]
    \centering
    \includegraphics[width=0.5\textwidth]{
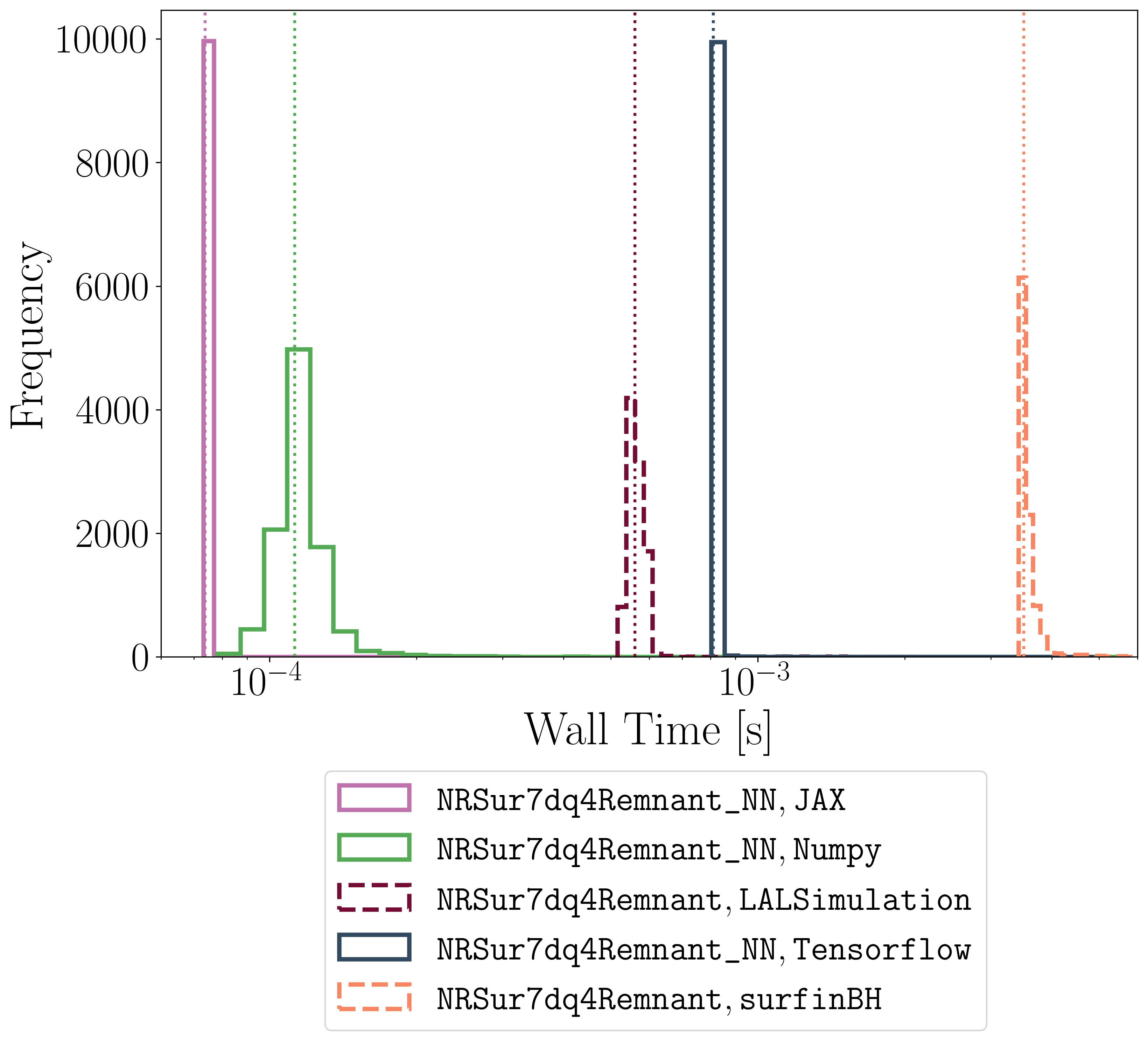}
    \caption{Evaluation timing comparison on a single CPU of \Remnant~(orange and purple) against \RemnantNN~(pink, green and blue), over the unseen \Sset. The dashed vertical lines show the median for each distribution.
    For each surrogate model, we further consider multiple implementations; see text for details. 
    We find a runtime speedup of around an order of magnitude with the \jax implementation, to a median of $7.4\times10^{-5}\,$s, while a \numpy implementation achieves a median time of $1.1\times10^{-4}\,$s, and the native \texttt{Tensorflow} implementation a median time of $8.1\times10^{-4}\,$s. The median evaluation times for the \Remnant~model are $3.5\times10^{-3}\,$s and $5.6\times10^{-4}\,$s for the \texttt{surfinBH} and \texttt{LALSimulation} implementations respectively.
    }
    \label{fig:cpu_timing}
\end{figure}

Having verified the accuracy of \RemnantNN, we now turn to its evaluation performance. 
We start with CPU timing tests using an Apple M2 chip with 8 cores.
In Fig.~\ref{fig:cpu_timing} we compare evaluation times over the \Sset~for the following implementations. 
For \RemnantNN, we consider the native \tensorflow~architecture and also a \numpy matrix implementation, where we take the final network weights and biases, and apply them to layers of a \numpy array, also using \numpy implementations of the relevant activation function in between. We additionally consider a \jax implementation, where similarly to the \numpy implementation, the weights and biases are applied to a \jax array with \jax activation functions.
For \Remnant, we consider the implementations in \texttt{surfinBH} \cite{surfinbh} and in \texttt{LALSimulation} \cite{lalsuite}.
Between the \Remnant~implementations, the one in \texttt{LALSimulation} is faster with a median evaluation time of $5.6\times10^{-4}\,$s compared to the \texttt{surfinBH} one at $3.5\times10^{-3}\,$s. 
We obtain a median evaluation time of $7.4\times10^{-5}\,$s with \jax, $1.1\times10^{-4}\,$s with \numpy and $8.1\times10^{-4}\,$s with \tensorflow.\footnote{\texttt{Tensorflow} evaluates slower than a simpler \jax or \numpy implementation without GPU parallelization due to the overhead costs of performing operations on tensor objects as opposed to arrays. Differences in framework construction may mean these overheads are lessened with other backends, for example \texttt{PyTorch}.}
Therefore, the neural network-based model represents a speed up of nearly an order of magnitude with the \jax implementation.

\begin{figure}[]
        \centering
    \includegraphics[width=0.49\textwidth]{
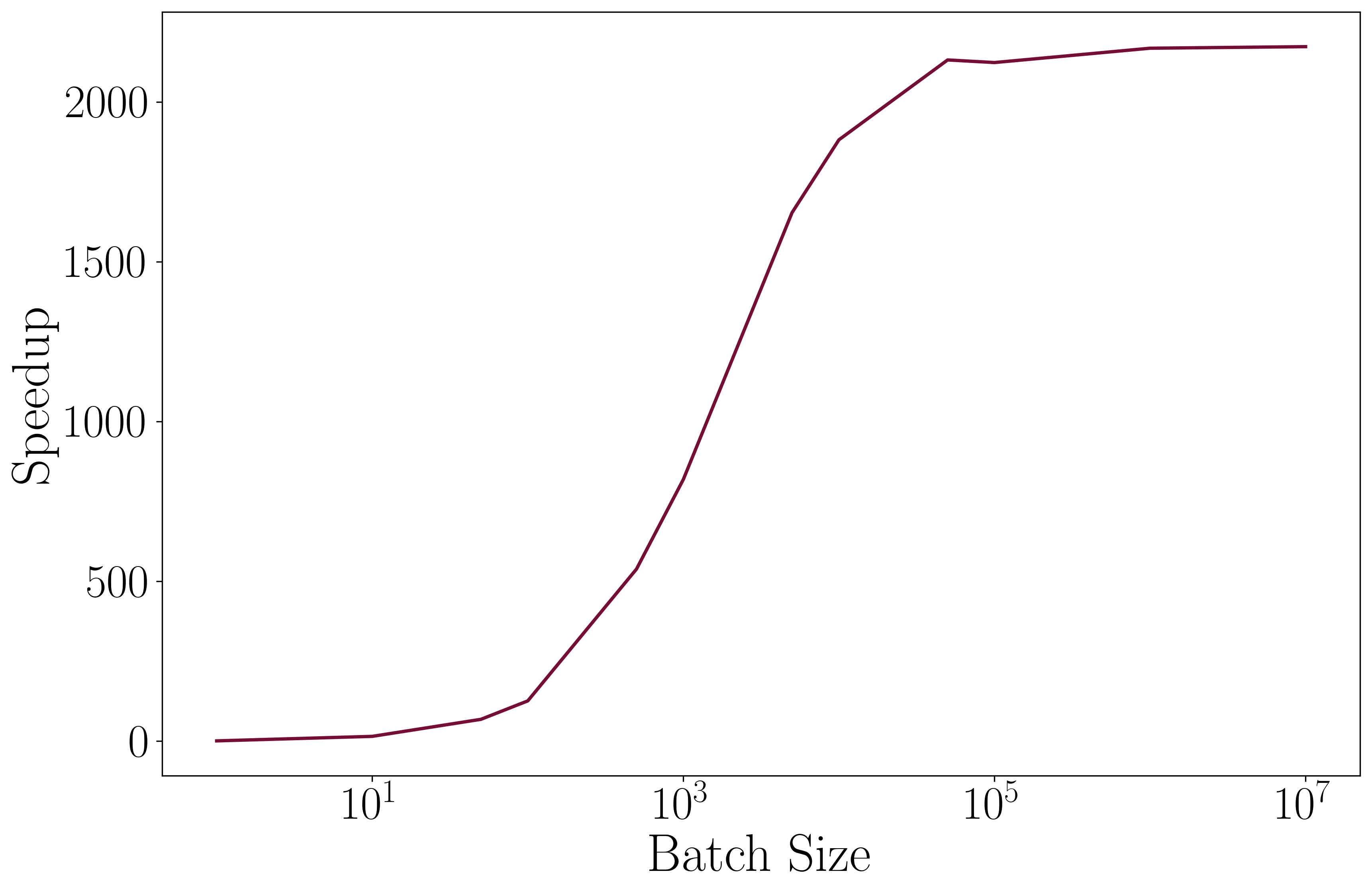}
    \caption{Evaluation timing on an NVIDIA A100-SXM4-40GB GPU of the \tensorflow~ \RemnantNN~implementation, comparing speedups as a function of batch size due to efficient parallel processing. 
    The speedup is defined as the batch size divided by the evaluation time.
    We find a significant speed up compared to single-point evaluation as the batch size increases up to $10^5$ points, after which the speed up plateaus.}
    \label{fig:gpu_timing}
\end{figure}

As the neural network is designed to be efficiently evaluated in batch, we also test the native \tensorflow~implementation as a function of batch size using an NVIDIA A100-SXM4-40GB GPU. 
We draw points uniformly from the training parameter space and evaluate the model in batch sizes up to $10^{7}$ points. 
Figure~\ref{fig:gpu_timing} shows the speedup, defined as the batch size divided by the evaluation time, as a function of batch size.
As the batch size increases, so does the speedup up to a batch size of $10^{5}$, after which all of the GPU processing units have been utilized.
Even larger batches must be evaluated sequentially in units of $10^{5}$, so the speed up remains constant. 
The maximum speed up from batch evaluation exceeds $3$ orders of magnitude.
Combined with the speed up just from the CPU evaluation, it leads to an overall potential maximum speed up of around $4-5$ orders of magnitude between the original surrogate model and its neural network version. 

\begin{figure}[]
    \centering
    \includegraphics[width=0.49\textwidth]{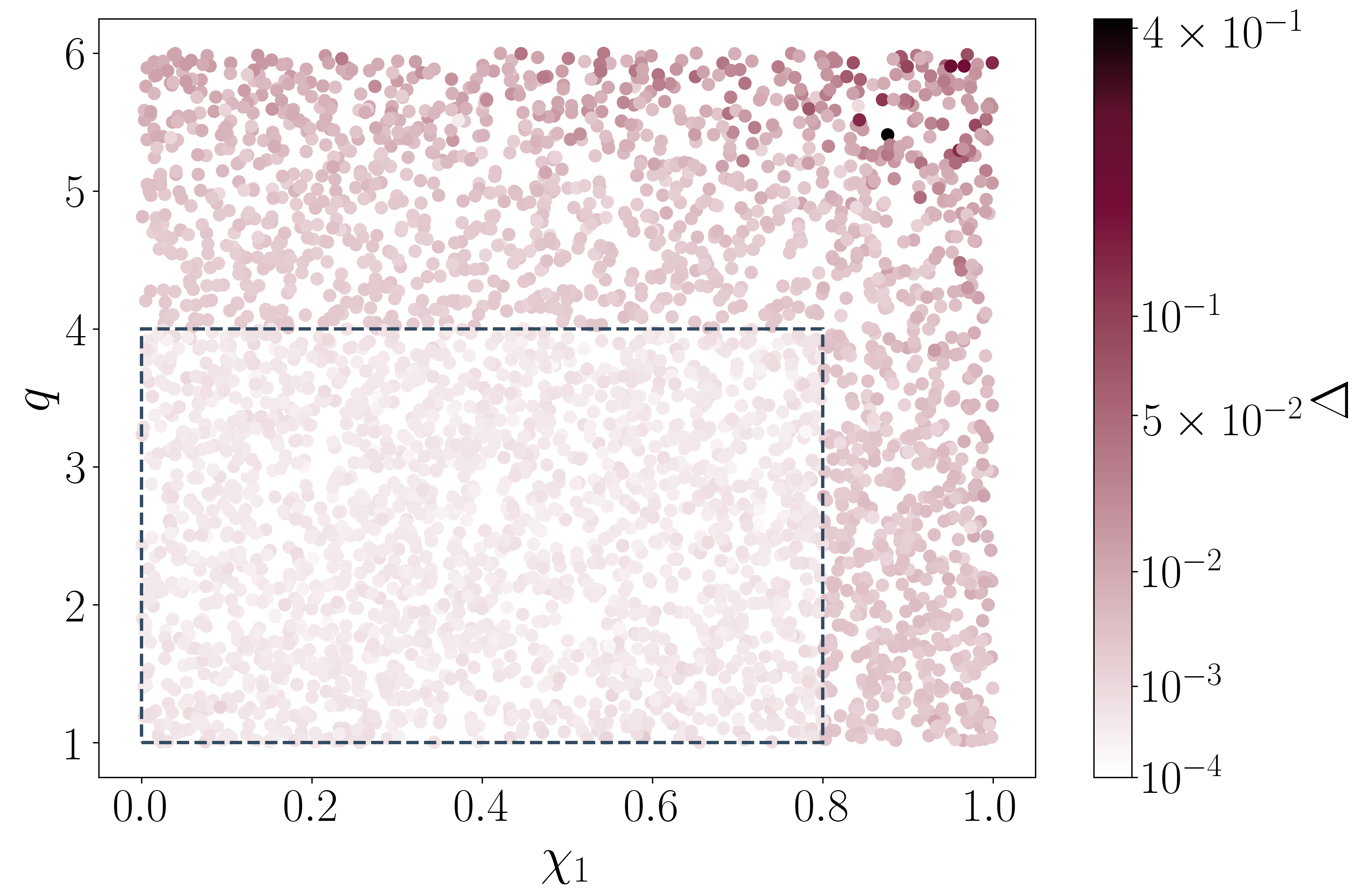}
    \caption{Random points on the $q-\chi_1$ plane colored by their mean error $\Delta$.
    The dashed rectangle denotes the training region, $q\in\left[1,2\right]$, $\chi_1\in[0,0.8]$.
    Points outside the rectangle are extrapolated at mass ratio $q\in\left[4,6\right]$ and primary spin magnitudes $\chi_1\in[0.8,1.0]$. 
    The mean error is higher across the extrapolation regions than across the training region. The spin magnitudes extrapolate more accurately than the mass ratio, and the worst-performing points have near-maximal spin magnitude and $q$ close to $6$.}
    \label{fig:accuracy_extrap}
    \end{figure}

Lastly, we test how the model extrapolates outside its parameter space training region.
Extrapolation from $q=4$ to $6$ was considered in Ref.~\cite{Varma:2019csw} and here we also consider spin.
In this test, the ground truth model remains \Remnant~rather than NR. 
This test is therefore assessing how \RemnantNN~extrapolates over \Remnant, rather than its accuracy in reproducing NR data in that domain.
We draw uniformly $1,000$ with only mass ratio extrapolation, $q\in\left[4,6\right]$, $500$ points with only primary spin magnitude extrapolation, $\chi_1\in\left[0.8,1.0\right]$, and a further $250$ points where both the mass ratio and primary spin magnitude are extrapolated, $q\in\left[4,6\right]$, $\chi_1\in\left[0.8,1.0\right]$.
For reference, we also consider $1,000$ points from the \Sset, namely from within the training region.
We evaluate these points with \RemnantNN~and plot them in the $q$ and $\chi_1$ space, colored by the mean error $\Delta$. The results are shown in Figure \ref{fig:accuracy_extrap}.
Across the training region (denoted by the dashed rectangle), the mean errors are fairly uniform and around $10^{3-4}$, consistent with Fig.~\ref{fig:FinalAccuracyResults}.
As the mass ratio and/or spin magnitude increases outside this region, the mean errors increase rapidly. 
Mean errors for extrapolated spin magnitudes remain fairly consistent as long as the mass ratio stays within its training region, never exceeding $4\times10^{-2}$. 
The extrapolated mass ratio region (with the spin remaining less than $0.8$) sees larger errors, with $q~{\sim}6$ resulting in errors up to $2\times10^{-1}$. 
The worst errors are for points where both the mass ratio and spin magnitude are extrapolated, with errors for maximal spins and mass ratios reaching up to $3\times10^{-1}$. 
These results suggest that extrapolation along mass ratio is more perilous than along spin magnitude and further NR simulations with unequal masses are required to improve surrogate models. 

\section{Conclusions}
\label{sec:Conclusions}

In this work, we have developed a systematic approach to optimizing the hyperparameters and training datasets of neural network surrogate models. 
We have applied our methodology to BBH remnant properties, building a new model, \RemnantNN, based on training data from the existing \Remnant~model. 
We achieve comparable accuracy with respect to the underlying model to the accuracy of \Remnant~to its underlying NR simulations for the remnant mass and final spin vector. 
The accuracy for the recoil velocity is about an order of magnitude higher.  

\RemnantNN~evaluates faster than the \texttt{surfinBH} implementation of \Remnant~on a single CPU when using the native \texttt{Tensorflow} architecture, with a median speedup of $4.3$ times, but performs slower than the \texttt{LALSimulation} implementation, by a factor of $1.4$.
We achieve greater speedups when using a \texttt{Numpy} implementation, improving to speedup factors of $5$ and $31$ compared to \texttt{LALSimulation} and \texttt{surfinBH} respectively. Our greatest speedups are obtained using a \jax implementation, by $8$ and $47$ times compared to \texttt{LALSimulation} and \texttt{surfinBH} respectively.
When evaluating in batch on a GPU, we obtain additional speedups up to $2000$ times, resulting in an overall maximum speedup of ${\sim}4$ orders of magnitude. 

These accuracies and speedups, however, come at the expense of an order of magnitude more training data than the available NR catalogs. 
We therefore propose such neural network-based surrogates as a secondary step after initially building an NR surrogate using different methods.
The neural network-based surrogate can then unlock significant speedups without compromising accuracy.
Our method is applicable to waveform surrogates as well as remnant properties, and we leave this application to future work. While we have used neural networks in a secondary NR surrogate-building stage, an alternative could be to use transfer learning from a neural network surrogate of a different model, for example an effective-one-body model, where training data is more readily available. In this alternative approach, the neural network surrogate would then be fine-tuned on the smaller pool of NR waveforms, and we leave investigations of this method to future work.

Such future work could also explore further improvements.
Firstly, we do not explore the impact of re-parameterization of the training data. 
Through initial investigations, we find that binary spins are more easily fitted when expressed in spherical rather than Cartesian coordinates. 
We do not explore further techniques, such as principle component analyses or dimensional reduction, which might be useful for waveform surrogates where the training data have many more dimensions. 
Secondly, we also do not consider the evolution of input parameters to different reference frequencies, instead performing our fit at the same reference time ($100\,$$M$ before merger) as with the original \Remnant~model. 
Thirdly, one limitation of the neural network is that it does not provide fitting error estimates in the same way as GPR fitting does. 
Therefore we can only assess model accuracy from the errors with respect to \Remnant~over a pre-defined test set.

Finally, the optimization procedure is not guaranteed to achieve a global minimum in the final loss, especially since we split optimization into distinct stages: functional hyperparameter, size and shape parameters, and training dataset. 
Additionally, because each sample neural network in the optimization is initialized using a single random seed, there is a chance that the globally optimal configuration underperforms with this seed and is discarded.
However, it will achieve a local minimum, which we assume to be sufficient (and indeed confirm so for \RemnantNN), and it is guaranteed to finish in a finite time. 
If global accuracy is required for future surrogate models, optimization could be done with several seed points in parallel, for a better chance of reaching the true global minimum loss.

\RemnantNN~is publicly available at~\cite{surfinbh} and \texttt{gwbonsai} at~\cite{gwbonsai}.

\section*{Acknowledgments}

LMT is supported by NSF MPS-Gravity Award 2207758 and by NSF Grant 2309200.
KC was supported by NSF Grant PHY-2409001.
V.V.~acknowledges support from NSF Grant No. PHY-2309301.
S.E.F.~acknowledges support from NSF Grants PHY-2110496 and AST-2407454.
S.E.F.~and V.V.~were supported by UMass Dartmouth's Marine and Undersea
Technology (MUST) research program funded by the Office of Naval Research 
(ONR) under grant no. N00014-23-1-2141.
The authors are grateful for computational resources provided by the LIGO Laboratory and supported by National Science Foundation Grants PHY-0757058 and PHY-0823459.

\bibliography{References}

\end{document}